\documentclass[apj]{emulateapj}
\slugcomment{{\sc Accepted to ApJ:} September 30, 2004}

\def\ni{\noindent}

\def\yr{{\rm\,yr}}

\def\pomega{\varpi}

\renewcommand{\v}[1]{\ensuremath{\mathbf{#1}}}
\newcommand{\beq}{\begin{equation}}
\newcommand{\eeq}{\end{equation}}
\newcommand{\ptl}{\partial}

\def\subN{_{\rm N}}

\usepackage{epsfig}
\usepackage{amsmath}

\begin{document}

\shortauthors{Murray-Clay \& Chiang}
\shorttitle{Origin of Asymmetric Capture}

\title{A Signature of Planetary Migration: The Origin of Asymmetric Capture in the 2:1 Resonance}

\author{Ruth A.~Murray-Clay\altaffilmark{1} \& Eugene I.~Chiang\altaffilmark{1,2}}

\altaffiltext{1}{Center for Integrative Planetary Sciences,
Astronomy Department,
University of California at Berkeley,
Berkeley, CA~94720, USA}
\altaffiltext{2}{Alfred P.~Sloan Research Fellow}

\email{rmurray@astron.berkeley.edu,\\ echiang@astron.berkeley.edu}

\begin{abstract}
The spatial distribution of Kuiper belt objects (KBOs) in 2:1 exterior resonance with Neptune constrains that planet's migration history.  Numerical simulations demonstrate that fast planetary migration generates a larger population of KBOs trailing rather than leading Neptune in orbital longitude.  This asymmetry corresponds to a greater proportion of objects caught into asymmetric resonance such that their resonance angles, $\phi$, librate about values $>\pi$ (trailing) as opposed to $<\pi$ (leading).  We provide, for the first time, an explanation of this phenomenon, using physical, analytic, and semi-analytic arguments.  Central to our understanding is how planetary migration shifts the equilibrium points of the superposed direct and indirect potentials.  Symmetric libration, in which $\phi$ librates about $\sim$$\pi$, precedes capture into asymmetric resonance.  As a particle transitions from symmetric to asymmetric libration, if $\phi$ exceeds its value, $\psi$, at the unstable point of asymmetric resonance, then the particle is caught into trailing resonance, while if $\phi<\psi$, the particle is caught into leading resonance.  The probability that the KBO is caught into trailing resonance is determined by the fraction of time it spends with $\phi>\psi$ while in symmetric libration. This fractional time increases with faster migration because migration not only shifts $\psi$ to values $<\pi$, but also shifts the stable point of symmetric libration to values $>\pi$.  Smaller eccentricities prior to capture strengthen the effect of these shifts.  Large capture asymmetries appear for exponential timescales of migration, $\tau$, shorter than $\sim$$10^7$ yr. The observed distribution of 2:1 KBOs (2 trailing and 7 leading) excludes $\tau\leq10^6$ yr with 99.65\% confidence.  
\end{abstract}


\section{INTRODUCTION}
\label{sec-intro}

Transfers of energy and angular momentum between disks and bodies embedded
within them can explain a variety of observed dynamical architectures.
For example, the tight orbits of
close-in, extra-solar, Jovian-mass planets (``hot Jupiters'') are thought
to be a consequence of orbital migration driven by parent disks
composed either of viscous gas (e.g., Ward 1997) or planetesimals
(e.g., Murray et al.~1998). Capture of bodies into mean-motion
resonances is a celebrated signature of migration (see, e.g., Peale 1986, and references therein),
showcased recently by the pair of planets orbiting GJ 876 in a 2:1 resonance
(Marcy et al.~2001; Lee \& Peale 2002). Migration finds application in the solar system as well; it promises to explain the preponderance of highly
eccentric Kuiper belt objects (KBOs) trapped in mean-motion resonances
with Neptune (Malhotra 1995; Chiang et al.~2003ab).
The outermost planet in our solar system might have migrated several AUs
outward by gravitationally scattering planetesimals
(Fernandez \& Ip 1984; Hahn \& Malhotra 1999), thereby sweeping
its exterior resonances across the primordial Kuiper belt and filling
them with KBOs (Malhotra 1995; Chiang \& Jordan 2002).

Notwithstanding the numerous appeals to disk-driven migration,
it has been difficult to dispel all doubt regarding its relevance.
Worries are perhaps most naggingly persistent in the Kuiper belt,
where the abundance of resonant KBOs could, in principle, be
explained by the following stability argument. One imagines that the entire
belt was once indiscriminately, dynamically excited to large eccentricities and
inclinations by one or more massive perturbers---perhaps Neptune
itself, during that planet's high eccentricity phase
(Goldreich, Lithwick, \& Sari 2004; see also Thommes, Duncan, \& Levison
1999, 2002)---and that only KBOs fortunate enough to be
kicked into resonances enjoyed the phase protection
that permits survival for the age of the solar system.

Chiang \& Jordan (2002, hereafter CJ) propose one test of the migration
hypothesis that helps to break degeneracies of interpretation.
They find that in scenarios where Neptune's
migration is comparatively fast---occurring on timescales shorter than 
$10^7$ yr, for their chosen initial
conditions---more objects are caught into the 2:1 resonance with mean libration angles
greater than $\pi$. This asymmetry in the distribution
of libration angles manifests itself as an asymmetry
on the sky: at a given
epoch, more 2:1 resonant objects will appear at longitudes trailing, rather than leading, Neptune's. This is a prediction
of the migration model that can be tested observationally
by wide-angle, astrometric surveys such as, e.g., Pan-STARRS
(Panoramic Survey Telescope \& Rapid Response System).  Under either the alternative stability hypothesis or the hypothesis that Neptune's migration occurred over timescales longer than $10^7$ yr, we would expect equal populations trailing and leading Neptune.

The capture asymmetry discovered by CJ---witnessed
also in numerical simulations by Wyatt (2003) in the context
of debris disks molded by migrating extra-solar planets---arose
from a purely numerical orbital integration.
In this paper, we explain the capture asymmetry on physical
grounds, using a variety of qualitative, analytic, and numerical
descriptions. In uncovering the workings of the capture
asymmetry on a step-by-step, mechanistic basis, we gain insight
into its applicability to observations. Our focus
is on the Kuiper belt, but the setting is clearly generalizable.

Terminology in this field can be confusing; we lay down
some definitions here. The exterior 2:1 resonance
is distinguished in offering the possibility of multiple
stable points for the resonance angle, 
\beq\label{eqn-phi}
\phi \equiv 2\lambda - \lambda_{\rm N}- \pomega \,\, , 
\eeq
where $\lambda$ and $\lambda_{\rm N}$ are the mean
longitudes of the test particle (KBO) and the planet (Neptune), respectively,
and $\pomega$ is the longitude of pericenter of the particle.\footnote{We label $p$:$q$ resonances such that $p>q$ refers to exterior resonances (outside the perturber) and $p<q$ refers to interior resonances (inside the perturber). This choice runs counter to convention, but seems commonplace in the Kuiper belt literature.} At
moderate eccentricities of the particle, the angle $\phi$ can
librate (oscillate with bounded amplitude) about a range of
values.  Particles
whose libration centers do not equal $\pi$
are said to be in ``asymmetric resonance'' or ``asymmetric
libration'' (Message 1958; Frangakis 1973; Beaug\'{e} 1994; Malhotra 1996; Winter \& Murray 1997; Pan \& Sari 2004, and references therein). The usual
libration of $\phi$ about $\pi$ is called ``symmetric resonance.''
We explain the physical origins of symmetric and asymmetric resonance
in \textsection\ref{sec-libration}.
We refer to the migration-induced preference
for objects to be caught into asymmetric resonance such that
they librate about angles greater than $\pi$
as ``asymmetric capture.''  Note that asymmetric capture does not refer merely to capture into asymmetric resonance.
We explain the mechanics underlying asymmetric capture
in \textsection\ref{sec-capture}. Connections to observations
are made in \textsection\ref{sec-observations}, and a summary and outlook
for future work are presented in \textsection\ref{sec-summary}.

\section{THE ORIGIN OF ASYMMETRIC LIBRATION}
\label{sec-libration}

We begin by explaining the origin of asymmetric libration in a static 2:1 resonant potential.  Our analysis of this special case of the circular, restricted, planar 3-body problem lays the foundation for understanding the more complicated situation in which the planet is migrating.  We offer two viewpoints: a physical, qualitative description in terms of impulses imparted to the particle over a synodic period (\textsection \ref{sec-potentials}), and a graphical, quantitative interpretation using contour plots of constant Hamiltonian (\textsection \ref{sec-contours}).

Many of the ideas contained in this section are not new; they may be traced in various forms in pioneering work on asymmetric resonance by Message (1958), Frangakis (1973), Beaug\'e (1994), Malhotra (1996), Winter \& Murray (1997), and Pan \& Sari (2004).  In particular, Frangakis (1973) and Pan \& Sari (2004) highlight the key role played by the indirect potential.  Pan \& Sari (2004) additionally offer physical explanations of asymmetric resonance, concentrating on the large eccentricity limit and on how impulsive torques imparted to the particle near its periapsis by the central mass and by the perturber can balance. We focus instead on the regime of small-to-moderate eccentricities.  Our methodology owes more to that of Peale (1986).  We proceed with our version of the facts in the spirit of pedagogy and to establish the language in which we will describe the dynamics of asymmetric resonance when the perturber migrates (\textsection\ref{sec-capture}), a subject which seems to have received much less attention.

\subsection{Physical Interpretation of the Direct and Indirect Potentials}
\label{sec-potentials}


Asymmetric libration results from the superposition of the direct and indirect disturbing potentials.  Consider the interaction between the Sun, Neptune, and a massless KBO in a reference frame centered on the Sun.  Denote the mass of the Sun by $m_{\odot}$, the mass of the planet by $m_{\rm N}$, the distance between the Sun and the planet by $\v{r_N}$, and the distance
between the Sun and the test particle by $\v{r}$. Then the
acceleration of the KBO equals
\beq 
\v{\ddot{r}} = \nabla_\v{r}(U_{\rm Kep} + R) \,\, ,
\label{eqn-accel}
\eeq
where
\beq
U_{\rm Kep} = \frac{Gm_\sun}{|\v{r}|}
\label{eqn-kep}
\eeq
is the Keplerian potential felt by the KBO, and
\beq
R = R_{\rm direct} + R_{\rm indirect} = \frac{Gm\subN}{|\v{r}-\v{r_N}|} - \frac{Gm\subN\v{r_N}\cdot\v{r}}{|\v{r_N}|^3} 
\label{eqn-disturb}
\eeq
is the disturbing or perturbation potential due to the planet.  Here $\nabla_\v{r}$ is the gradient with respect to \v{r}.  The perturbation potential divides into a direct part, $Gm\subN/|\v{r}-\v{r_N}|$, which governs the direct attraction between the planet and the test particle, and an indirect part, $-Gm\subN\v{r_N}\cdot\v{r}/|\v{r_N}|^3$, which accounts for the acceleration of the reference frame due to Neptune's acceleration of the Sun.  Note that in an arbitrary accelerating reference frame, there is no guarantee that the acceleration of a body can be written as the gradient of a potential.  We are able to do so because the acceleration of our reference frame corresponds to the gravitational acceleration of the Sun by Neptune.

We describe the evolution of the test particle's orbit inside the exterior 2:1 resonance in terms of its resonance angle, $\phi = 2\lambda-\lambda\subN-\pomega$.  The angle $\phi$ is approximately the angle of the line drawn through the planet and the particle at conjunction, measured from pericenter. This geometric interpretation is only approximate because $\phi$ is written in terms of mean longitudes rather than true longitudes.

\begin{figure}
\epsscale{1.1}
\plotone{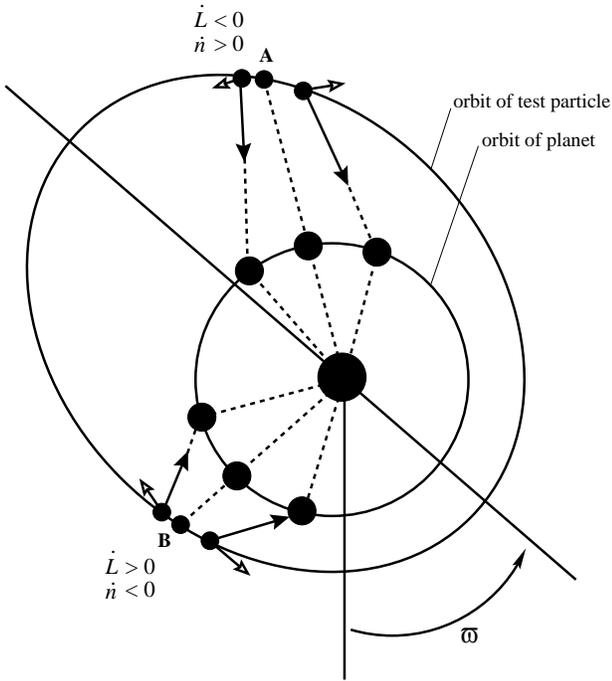}
\caption{Schematic diagram of the effect of the direct disturbing potential when conjunctions occur at points A and B (not to scale).  Solid arrows represent direct accelerations of the particle by the planet.  Open arrows are the azimuthal components of these accelerations.  The angular momentum and mean motion of the test particle are $L$ and $n$, respectively.  Conjunctions at point A ($0<\phi<\pi$) remove angular momentum from the test particle, increase its mean motion, and accelerate $\phi$.  Conjunctions at point B result in opposite behavior.  After figure 1 of Peale (1986).}
\label{fig-direct}
\end{figure}

\placefigure{fig-direct}

Peale (1986) describes how $\phi$ evolves under the direct perturbation, and we summarize his analysis here.  From conjunction to opposition, the direct perturbation adds angular momentum to the KBO through the azimuthal acceleration exerted by Neptune.  From opposition to the next conjunction, the azimuthal acceleration removes angular momentum.  Over a synodic period, the sign of the effect of the direct perturbation on $\phi$ can be decided by examining interactions near conjunction, when Neptune and the KBO are closest.  Consider conjunctions at point A of Figure \ref{fig-direct}, en route from pericenter to apocenter ($0<\phi<\pi$).  The angular momentum removed from the KBO before conjunction exceeds the angular momentum imparted after conjunction for two reasons.  First, the two orbits are diverging at point A. Second, the difference in angular velocities of Neptune and of the KBO is smaller before conjunction than after, so that the two bodies spend more time close to each other before conjunction.  The net effect of the conjunction is therefore to pull the KBO backwards along its orbit, removing its orbital angular momentum.  Its semi-major axis decreases, and its mean motion increases.\footnote{This argument neglects changes in the semi-major axis of the particle due to the radial component of the direct acceleration; these changes are smaller than those brought about by the azimuthal acceleration by of order the eccentricity of the particle.}  As a result, the next conjunction occurs later than it would if the planet and the test particle were non-interacting; in other words, if $0<\phi<\pi$, the direct potential increases $\phi$ over its value in the absence of the direct perturbation.  We can say that $\phi$ is accelerated to larger values.  Analogous considerations apply to conjunctions at point B in Figure \ref{fig-direct} ($\pi<\phi<2\pi$); Neptune adds angular momentum to the orbit of the KBO and $\phi$ decelerates.

The tendency of the direct perturbation to restore $\phi$ towards $\pi$ is analogous to the effect of a gravitational field on the motion of a pendulum, with $\phi=\pi$  and $\phi=0$ corresponding to the stable and unstable equilibrium points, respectively. 

\begin{figure}
\epsscale{1.2}
\plotone{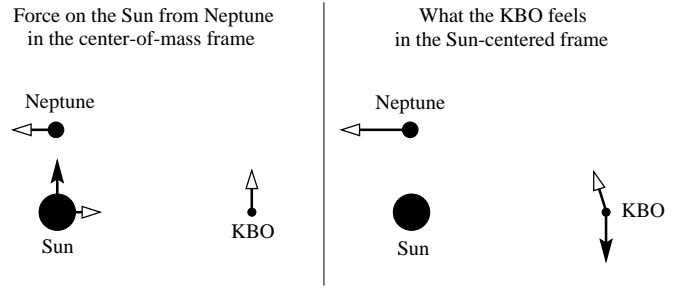}
\caption{The effect of the indirect potential.  Open arrows indicate the directions that bodies are moving in their orbits.  Solid arrows indicate accelerations. When Neptune accelerates the Sun in an inertial reference frame (left), the KBO feels a fictitious acceleration in the opposite direction in the Sun-centered frame (right).  Here Neptune is shown ahead of the KBO; in the Sun-centered frame, the indirect torque removes angular momentum from the KBO.}
\label{fig-indirect}
\end{figure}

\placefigure{fig-indirect}

\begin{figure}
\begin{center}
\epsscale{0.9}
\plotone{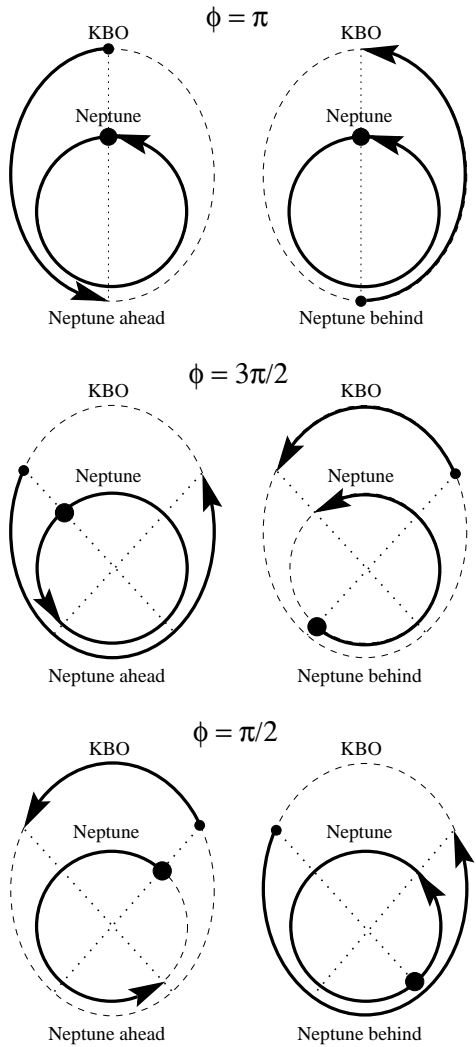}
\caption{Portions of the synodic period with Neptune leading and following the KBO for $\phi=\pi$, $\pi<\phi<2\pi$, and $0<\phi<\pi$ (top, middle, and bottom, respectively).  Solid circles indicate starting positions and arrow tips indicate ending positions.  For example, if $\phi=3\pi/2$, Neptune spends more time ahead of rather than behind the particle.}
\label{fig-indirect21}
\end{center}
\end{figure}

To understand asymmetric libration, we extend Peale's qualitative description to encompass the effect of the indirect perturbation.  As is true for the direct acceleration, the azimuthal component dominates the evolution of $\phi$.  Consider an infinite line connecting the Sun and the test particle.  We refer to the side of this line toward which the test particle is moving as ``ahead'' of the KBO and to the opposite side of the line as ``behind.'' From conjunction to opposition, Neptune is ahead of the KBO and therefore pulls the Sun ahead of the KBO.  As a result, in the Sun-centered frame, the KBO feels a fictitious torque which removes angular momentum from its orbit (Figure \ref{fig-indirect}).  Likewise, from opposition to the next conjunction, the fictitious indirect torque adds angular momentum to the KBO's orbit.  Integrated over the synodic period, the azimuthal component of the indirect acceleration equals
\beq
\langle T\rangle = -\frac{Gm\subN}{a\subN^2} \oint \sin\Delta\theta \,dt \,\, ,
\eeq
where $\Delta\theta \equiv \lambda\subN -f-\pomega$ is the angle between the true longitude, $\lambda\subN$, of the planet and the true longitude, $f+\pomega$, of the particle.  For $0<\phi<\pi$, Neptune spends more time behind the KBO, and $\langle T\rangle$ is positive;\footnote{The time spent by Neptune forward and backward of the KBO does not alone determine the sign of $\langle T\rangle$; one must examine the exact shape of $\sin\Delta\theta$ over a synodic period.  However, for the 2:1 resonance, the shape does not alter the sign of our argument.  We show in \textsection\ref{sec-summary} that $\langle T\rangle = -(Gm\subN/a\subN^2)\pi c_1$ for the 2:1 resonance; $c_1<0$ when $0<\phi<\pi$, and $c_1>0$ when $\pi<\phi<2\pi$.} 
the angular momentum of the KBO in the Sun-centered frame increases, and $\phi$ decelerates (Figure \ref{fig-indirect21}).  For $\pi<\phi<2\pi$, $\phi$ accelerates.  Thus, the indirect potential restores $\phi$ toward $0$. 

\placefigure{fig-indirect21}

Whereas $\phi$ responds to the direct perturbation like a pendulum, it responds to the indirect perturbation like a metronome.  The direct and indirect accelerations oppose one another; they can be balanced, forming stable points at values of $\phi$ intermediate between $0$ and $\pi$, and intermediate between $\pi$ and $2\pi$.  This balance underlies the phenomenon of asymmetric libration. 

In contrast, for the 3:2 resonance, Neptune spends the same amount of time behind and ahead of the KBO regardless of the value of its (appropriately defined) resonance angle, and the indirect torque averages to zero (see \textsection\ref{sec-summary} for a proof).  
Consequently, asymmetric libration is impossible for the 3:2 resonance.

\subsection{Contours of Constant Hamiltonian}
\label{sec-contours}

The long-term evolution of the KBO's orbit when Neptune's orbit is not varying can be summarized neatly using a Hamiltonian formulation.\footnote{We are allowed to use Hamilton's equations with the potential $V = U_{\rm Kep} + R$ because we can write all of the fictitious forces felt in our accelerating frame as gradients of fictitious potentials that depend only on the canonical coordinates, not the canonical momenta.  As long as this is the case, we can write the Lagrangian $\mathcal{L}=K-V$, where $K$ is the kinetic energy and $V$ includes the fictitious potentials.}  
We seek to write down a Hamiltonian having no explicit time-dependence so that we can plot level diagrams of the Hamiltonian and trace the evolution of $\phi$ graphically.  We will make these plots in two ways, using exact (\textsection\ref{sec-exact}) and series-expanded (\textsection\ref{sec-expanded}) versions of the Hamiltonian.

\subsubsection{Exact Hamiltonian}
\label{sec-exact}

We employ the Poincar\'e coordinates,
\beq
\lambda = M+\pomega, \;\;\; \gamma  = -\pomega,
\eeq
and corresponding momenta,
\beq
\Lambda = \sqrt{\mu a}, \;\;\; \Gamma = \sqrt{\mu a}(1-\sqrt{1-e^2}),
\eeq
where $\lambda$ is the mean longitude, $M$ is the mean anomaly, $\pomega$ is the longitude of pericenter, $a$ is the semi-major axis, and $e$ is the eccentricity, all appropriate to the KBO. Furthermore, $\mu\equiv Gm_\sun$.  These variables are modified from the standard set of Poincar\'e variables in order to describe the motion of a massless test particle. The Hamiltonian reads
\beq 
H = -\frac{\mu^2}{2\Lambda^2} - R(t) \,\, .
\eeq
The perturbation $R$ is a function of $\lambda\subN(t)$, which is explicitly time-dependent.

To eliminate the explicit time-dependence in $R$, we perform a point transformation.  Inside the 2:1 resonance, it is illuminating to employ the following coordinates:
\beq
\phi=2\lambda-\lambda\subN(t)-\pomega, \;\;\; \sigma=\lambda-\lambda\subN(t),
\eeq
for which the corresponding momenta equal
\beq
\Gamma, \;\;\; N = \Lambda-2\Gamma \,\, .
\eeq
We are interested in the evolution of $\phi$ (see \textsection\ref{sec-potentials}), and our choice for $\sigma$ will allow us to average over the synodic period easily.

The new time-independent Hamiltonian reads
\begin{align}\label{eqn-hamfull}
\widetilde{H} &= H-n\subN(\Gamma+N) \nonumber \\
&= -\frac{\mu^2}{2(N+2\Gamma)^2}-n\subN(\Gamma+N)-R(\phi,\sigma,\Gamma,N) \,\, ,
\end{align}
where $n\subN$ is the mean motion of Neptune.  The price we pay in eliminating the explicit time-dependence is the appearance of an extra term in the Hamiltonian, $-n\subN(\Gamma+N)=-n\subN|\v{r}\times\v{\dot r}|$. This term renders $\widetilde{H}$ equivalent to the Jacobi constant.  This formulation of the Hamiltonian will be useful when we allow Neptune to migrate in \textsection\ref{sec-capture}.

We are interested in the behavior of our system over timescales longer than the synodic period.  We therefore average over the synodic period to obtain our final Hamiltonian.  The averaged disturbing function equals
\begin{eqnarray} 
R_{\rm avg} &=&\frac{1}{2\pi}\int_0^{2\pi}R\;d\sigma \nonumber \\
&=& \frac{Gm\subN}{2\pi}\int_0^{2\pi}\left(\frac{1}{|\v{r}-\v{r\subN}|}-\frac{\v{r\subN}\cdot\v{r}}{|\v{r\subN}|^3}\right)d\sigma \,\, .
\end{eqnarray}
The total averaged Hamiltonian is
\beq 
H_{\rm avg}=-\frac{\mu^2}{2(N+2\Gamma)^2}-n\subN(\Gamma+N)-R_{\rm avg}(\phi,\Gamma,N) \,\, .
\label{eqn-hamavg}
\eeq
This Hamiltonian does not depend on $\sigma$.  Therefore $N$ is a constant of the motion: $\ptl H_{\rm avg}/\ptl\sigma=-\dot N=0$. 

\begin{figure}
\epsscale{1.4}
\vspace{-0.2in}
\hbox{\hspace{-0.3in}\plotone{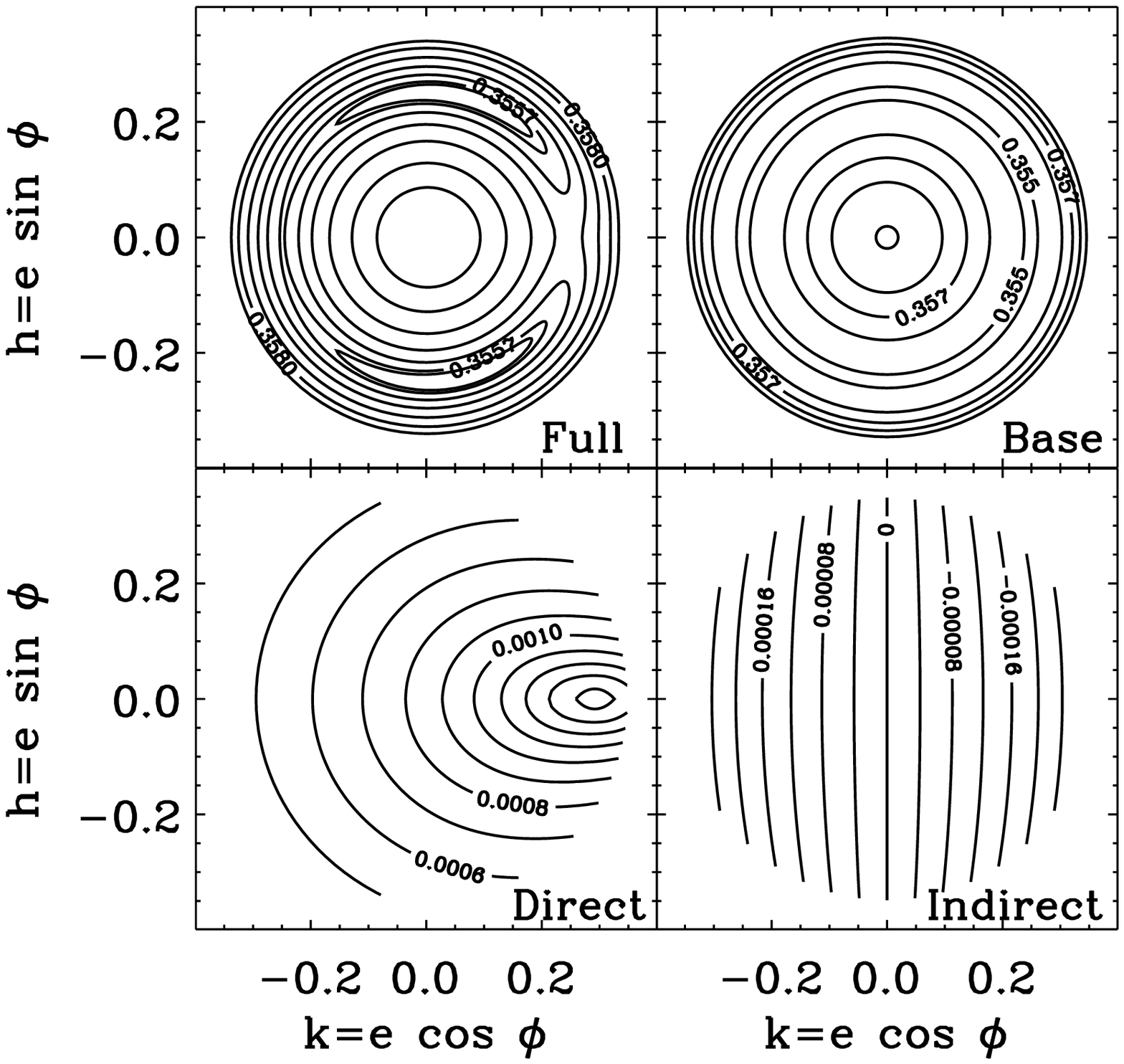}}
\caption{Contours of constant Hamiltonian for the exact Hamiltonian ($H_{\rm avg}$) and its components.  The contour labels give the value of $-H$ in the units $Gm_\sun =1$ and $a\subN = 1$. The total Hamiltonian, Keplerian plus coordinate-transformation terms (``base''), direct perturbation, and indirect perturbation are individually identified.  For this figure, $N=1.18\sqrt{Gm_\sun a\subN}$ and $m\subN = 10^{-3}m_\sun$.  Note that half of the contours for the indirect potential correspond to negative values of $R_{\rm indirect}$.}
\label{fig-components}
\end{figure}

\placefigure{fig-components}

Figure \ref{fig-components} displays level curves of $H_{\rm avg}$ as well as individual contributions to $H_{\rm avg}$ from the direct and indirect terms.  In Figures \ref{fig-components}, \ref{fig-orders}, and \ref{fig-sequence}, we set $m\subN = 10^{-3}m_\sun$.  Had we used Neptune's true mass ($m\subN = 5.15\times10^{-5} m_\sun$), contours of symmetric and asymmetric libration would be less pronounced.  When the planet is not migrating, the test particle's orbit evolves along a contour of constant Hamiltonian. The axes are $k=e\cos\phi$ and $h=e\sin\phi$, so that the distance from the center of each plot is the eccentricity and the azimuthal angle is $\phi$.  The semi-major axis and eccentricity of the test particle are related by
\beq\label{eqn-ae}
a=\frac{1}{\mu}\left(\frac{N}{2\sqrt{1-e^2}-1}\right)^2
\eeq
for a given $N$.  As $e$ increases, $a$ increases.  Thus, the distance from the center of each plot also represents the semi-major axis of the KBO.

The ``base'' Hamiltonian refers to the combination of the Kepler and coordinate transformation contributions:
\beq
H_{\rm base} = -\frac{\mu^2}{2(N+2\Gamma)^2}-n\subN(\Gamma+N) \,\, .
\eeq
The contours of constant $H_{\rm base}$ do not depend on the angular position of the planet or of the test particle.  The Kepler and coordinate-transformation contributions are each negative.  The absolute value of the Kepler contribution decreases as $a$ increases, and the absolute value of the coordinate-transformation contribution increases with $a$.  For a given adiabatic invariant $N$, $|H_{\rm base}|$ contains a local trough.

The direct and indirect contributions are much smaller in magnitude than the base Hamiltonian and thus are not important in most regions of $h$-$k$ space.  However, in the trough in $H_{\rm base}$ (near exact resonance), the gradient in $H_{\rm base}$ is small enough that the disturbing terms significantly alter the shapes of the contours.  
The direct term ``fills in'' the trough near $\phi=0$, leading to libration about $\phi=\pi$.
For appropriate values of $N$, the opposing contours from the direct and indirect perturbations superimpose on the trough in $H_{\rm base}$ to generate contours of asymmetric libration: two unstable points appear at $\phi=\pi$ and $\phi=0$, and two stable points appear at $0<\phi<\pi$ and $\pi<\phi<2\pi$.

\subsubsection{Series-Expanded Hamiltonian}
\label{sec-expanded}

\begin{figure}
\vspace{-0.2in}
\epsscale{1.4}
\hbox{\hspace{-0.2in}\plotone{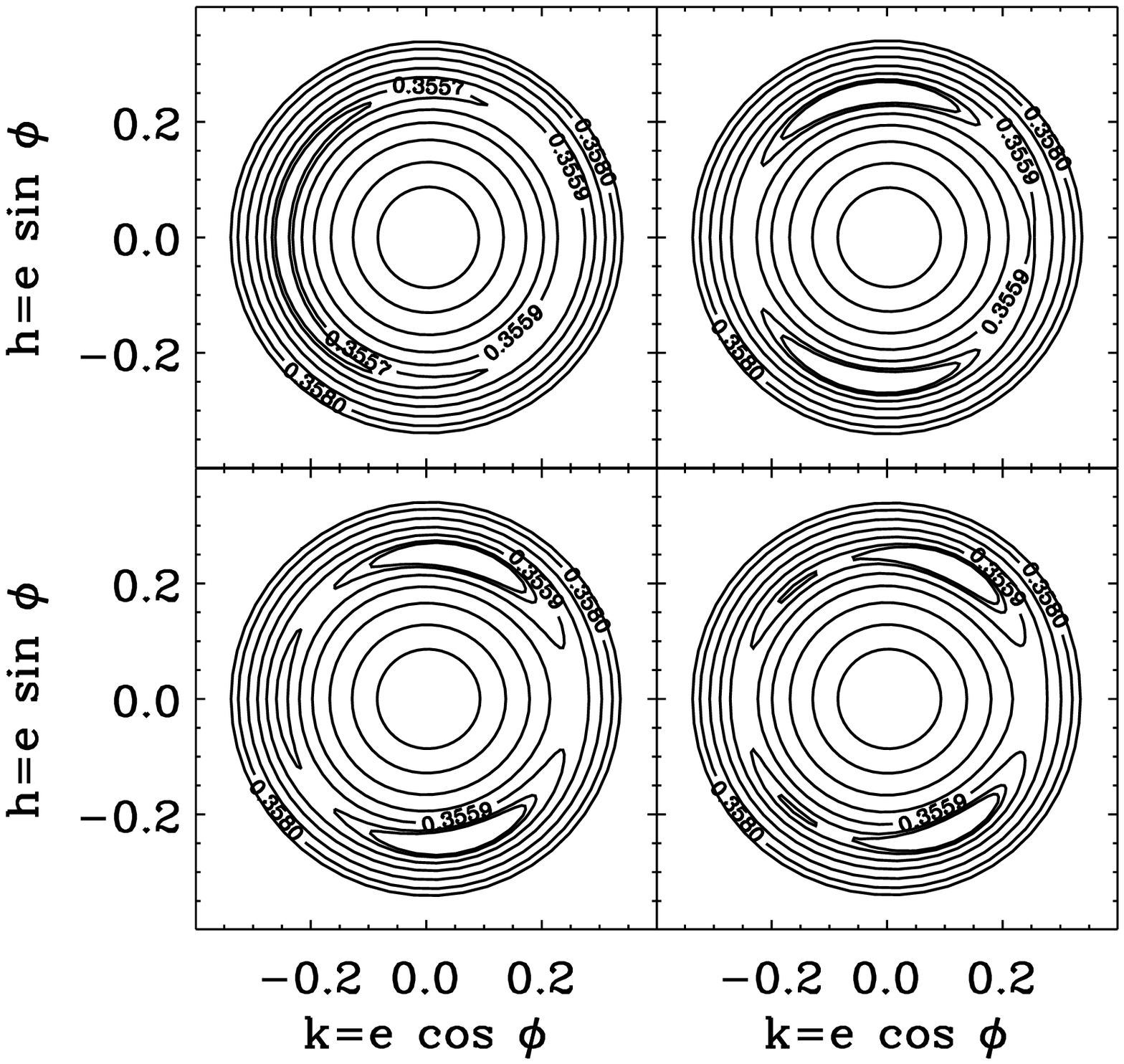}}
\caption{Contours of constant Hamiltonian with $R$ expanded to first, second, third, and fourth orders in $e$ (top left, top right, bottom left, and bottom right, respectively) for the same values of $N$, $a\subN$, and $m\subN$ used for Figure \ref{fig-components}.  The contour labels give the value of $-H_{\rm avg}$ in the units $Gm_\sun =1$ and $a\subN = 1$.  The second-order expansion compares favorably with the exact Hamiltonian (Figure \ref{fig-components}, top left panel), with small differences in the values of $\phi$ for the stable points.  In contrast, the third and fourth-order expansions poorly represent the exact Hamiltonian.}
\label{fig-orders}
\end{figure}

\placefigure{fig-orders}

We now approximate the Hamiltonian using a literal series expansion of the disturbing function, as presented in Murray \& Dermott (1999).  We must be careful about which terms we include.  Because the asymmetric islands are generated by small changes in the Hamiltonian, their shapes are sensitive to artifacts of the expansion.  Figure \ref{fig-orders} portrays the same Hamiltonian featured in Figure \ref{fig-components} using an expansion to first, second, third, and fourth orders in $e$.  The terms that depend on $\phi$ in this expansion are proportional to $\cos(j\phi)$, where $j$ is an integer ranging from $0$ to the order in $e$.  To first order, only a symmetric island can be represented.  To second order, we obtain a reasonable representation of asymmetric islands.  To third and fourth orders, more islands appear that are not present in the exact solution.  These artifacts do not appear for all values of $N$ and $a\subN$.  However, when the planet migrates (see \textsection\ref{sec-capture}), $N$ remains constant while $a\subN$ traverses a range of values.  At least some of the resulting combinations of $N$ and $a\subN$ exhibit artifacts if the expansion order is three or four.

To avoid this problem, we use the second-order expansion of $R$ in the integrations that follow:
\begin{align}\label{eqn-expand}
R_{\rm avg,2} =& \frac{Gm\subN}{a\subN}\left[\alpha(f_1+f_2e^2+f_{31}e\cos\phi + f_{53}e^2\cos 2\phi)- \nonumber \vphantom{\frac{1}{2\alpha}e\cos\phi}\right. \\
& \qquad\quad \left. \frac{1}{2\alpha}e\cos\phi\right] 
\end{align}
where $\alpha=a\subN/a$ and the $f_i$'s are functions of Laplace coefficients given in Murray \& Dermott (1999).  We have dropped terms that average to zero over the synodic period.  The Hamiltonian, expanded to second order and averaged over the synodic period, is
\beq\label{eqn-expandH}
H_{\rm avg,2} = H_{\rm base} - R_{\rm avg,2} \,\, .
\eeq
This second-order expansion generates contours of asymmetric libration but no artificial islands in inappropriate locations; it provides a good qualitative representation of the actual potential.  For example, the contours in the top left panel of Figure \ref{fig-components} and the top right panel of Figure \ref{fig-orders} are qualitatively similar, though the stable points in the two plots have somewhat different values of $\phi$.

Whereas Equation \ref{eqn-expand} suggests that asymmetric
libration arises from the second-order term in a power-series
expansion of the disturbing function, we emphasize that such
a conclusion is misleading from a physical point of view.
Asymmetric libration relies on a balance between the direct
and indirect perturbations; it depends on the indirect
term not time-averaging to zero. Expanding the direct term 
to infinite order while excluding the indirect term would not yield
asymmetric libration.  Beaug\'{e} (1994)
illustrates the difficulty of attributing asymmetric libration
to the second-order term by truncating the potential for the
3:2 resonance at second order and finding asymmetric libration
where there should be none. Equation \ref{eqn-expand} should be regarded
as simply a useful fitting formula for the true potential.

\section{THE ORIGIN OF ASYMMETRIC CAPTURE}
\label{sec-capture}

We now explain the origin of asymmetric capture when the planet is migrating.  As the planet migrates outward, it may capture test particles that begin outside resonance into resonance [see, e.g., Murray \& Dermott (1999), and references
therein].
To be caught into asymmetric resonance from an initially circulating orbit, a test particle must first be caught into symmetric resonance.  This may be seen qualitatively using a series of level diagrams for the Hamiltonian, and quantitatively by numerically integrating the equations of motion (\textsection\ref{sec-symasym}).  We recall from Goldreich (1965) that migration induces an offset in the stable point of symmetric libration (\textsection\ref{sec-offset}).  We expand upon these two observations to explain the differing probabilities of capture into the two islands of asymmetric resonance.  We elucidate three separate physical effects and provide a series of diagnostic numerical integrations (\textsection\ref{sec-decide}).  We provide a sample prediction of the analytic theory (\S\ref{sec-quant}), and compare it with numerical experiments that solve for the ratio of particles captured into libration about the two islands as a function of migration timescale and other initial conditions (\textsection\ref{sec-ratios}).

\subsection{The Transition from Circulation to Asymmetric Libration}
\label{sec-symasym}

\begin{figure}
\epsscale{1.4}
\vspace{-0.2in}
\hbox{\hspace{-0.2in}\plotone{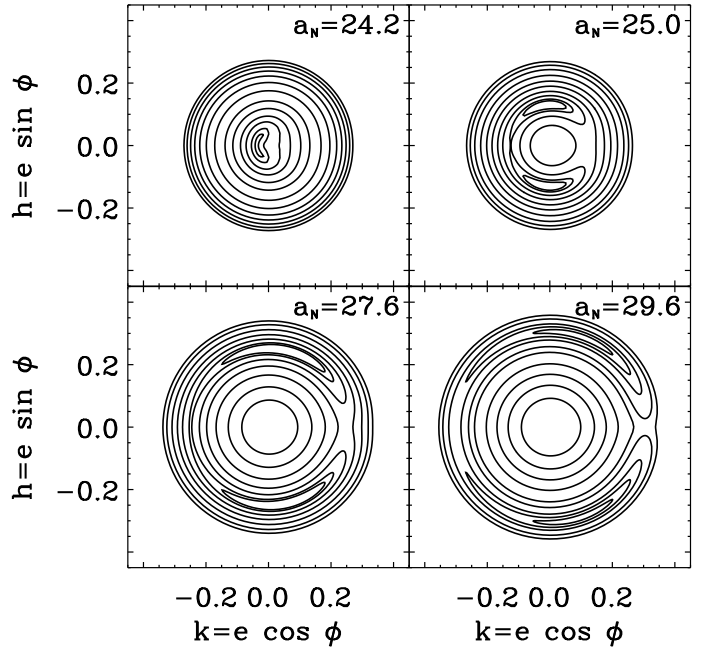}}
\caption{Sequence of level diagrams of $H_{\rm avg}$ for fixed $N=6.2\sqrt{Gm_\sun}\,{\rm AU}^{1/2}$ and increasing $a\subN$, marked on the plots in AU. Here, $m\subN = 10^{-3}m_\sun$. A particle whose $a$ is too large to be in resonance stays on approximately the same circulating contour at a large distance from the origin before resonance capture.  As Neptune migrates outward, the resonance expands outward to encounter the particle.}
\label{fig-sequence}
\end{figure}

Before a test particle can be captured into asymmetric libration, it must first be captured into symmetric libration.  To explain why this is so, we appeal to the contour plots discussed in \textsection\ref{sec-contours}.  Consider the limit in which the timescale for migration is long compared to the libration period.  Then, over timescales short compared to the migration time, the particle's orbit evolves approximately along a contour of constant exact Hamiltonian, $H_{\rm avg}$.\footnote{Level diagrams in this paper are drawn for fixed values of $a\subN$.  Level diagrams showing the exact paths of particles in $e$-$\phi$ space when $a\subN$ varies with time cannot be made because then the potential is time-dependent.}   Over timescales comparable to the migration time, we obtain a qualitative sense of the evolution of the particle's orbit by considering a progression of contour plots corresponding to different values of $a\subN$.  Such a progression is displayed in Figure \ref{fig-sequence}, in which panels correspond to the same value of the adiabatic invariant, $N$, but increasing values of $a\subN$.  Even when Neptune is migrating, the Hamiltonian (averaged over the synodic period) is still given by $H_{\rm avg}$; while $n\subN$ and $R_{\rm avg}$ are now explicit functions of time, $N$ is still a constant of the motion.

\placefigure{fig-sequence}

In the panels of Figure \ref{fig-sequence}, a larger value for $e$ corresponds to a larger value for $a$ by conservation of $N$ (Eqn. \ref{eqn-ae}).  As Neptune migrates outward, the libration centers migrate outward as well.  A particle having too large an $a$ to be in resonance circulates in the outer regions of the plots, maintaining approximately its same distance from the origin prior to resonance encounter.  The resonance appears to approach such a particle, which will eventually find itself on a contour of symmetric libration.  Since the asymmetric contours are all surrounded by symmetric contours and the evolution of the potential is smooth, a particle in asymmetric resonance must previously have occupied a symmetric contour.

We confirm the evolutionary sequence from circulation to symmetric libration to asymmetric libration by numerically integrating the equations of motion for $\phi$.  We employ the expanded Hamiltonian, $H_{\rm avg,2}$.  As discussed in \textsection\ref{sec-expanded}, this expansion reproduces the asymmetric islands in the exact Hamiltonian qualitatively well.  Our Hamiltonian equals
\begin{align} \label{eqn-hamilt}
&H_{\rm avg,2}(\phi,\Gamma,N,t) = \nonumber \\
&\qquad -\frac{\mu^2}{2(N+2\Gamma)^2}-n\subN(\Gamma+N)- \frac{Gm\subN}{a\subN}\left[\alpha(f_1+f_2e^2+ \vphantom{\frac{1}{2\alpha}e\cos\phi}\right. \nonumber \\
&\qquad\quad \left. f_{31}e\cos\phi + f_{53}e^2\cos 2\phi)-\frac{1}{2\alpha}e\cos\phi\right] \,\,
\end{align}
where $e=e(\Gamma,N)$.  We prescribe
\beq\label{eqn-migration}
a\subN = a_{{\rm N},0}\left(\frac{t}{t_0}\right)^{2/3} \,\, ,
\eeq 
where $a_{{\rm N},0}$ and $t_0$ are constants, so that $\alpha=\alpha(\Gamma,N,t)$ and $n\subN = n\subN(t)$.  The $f$'s are held constant in our computations.  The form of this prescription for Neptune's migration will be motivated in the next section.
  
\begin{figure}
\epsscale{1.4}
\hbox{\hspace{-0.5in}\plotone{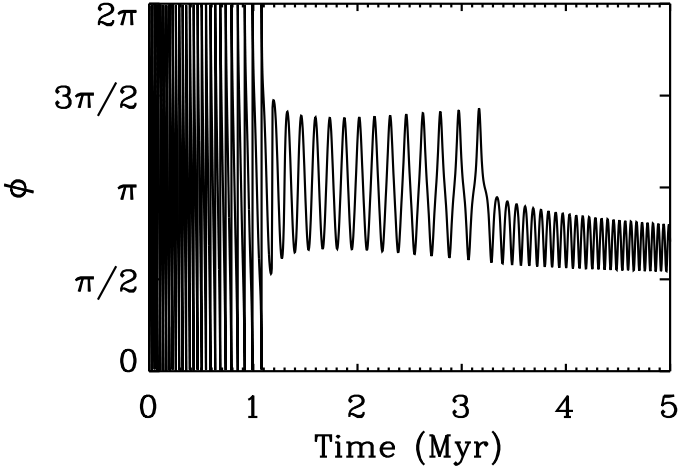}}
\caption{Evolution of $\phi$ from circulation to symmetric libration to asymmetric libration as the planet migrates outward.  In this figure, $m\subN$ is set equal to Neptune's true mass. The planet migrates according to Equation \ref{eqn-migration} with $t_0 = 300$ Myr and $a_{{\rm N},0} = 23.1$ AU.  Time on the plot is measured from $t=311$ Myr when the particle is characterized by $\phi=0$, $e=0.01$, and $a=37.67$ AU (1 AU outside of nominal resonance).}
\label{fig-threephase}
\end{figure}

The equations of motion,
\beq
\dot\phi = \frac{\ptl H_{\rm avg,2}}{\ptl\Gamma}, \;\;\; \dot\Gamma = -\frac{\ptl H_{\rm avg,2}}{\ptl\phi} \,\, ,
\eeq
are integrated using the Bulirsch-Stoer algorithm (Press et~al.~1992).
A sample time evolution for $\phi$ is showcased in Figure \ref{fig-threephase}, with input parameters listed in the caption.  The particle initially circulates, then librates symmetrically, and finally is captured into asymmetric libration.

\placefigure{fig-threephase}

\subsection{Migration-Induced Offset in Symmetric Libration}
\label{sec-offset}

\begin{figure}
\epsscale{1.4}
\hbox{\hspace{-0.5in}\plotone{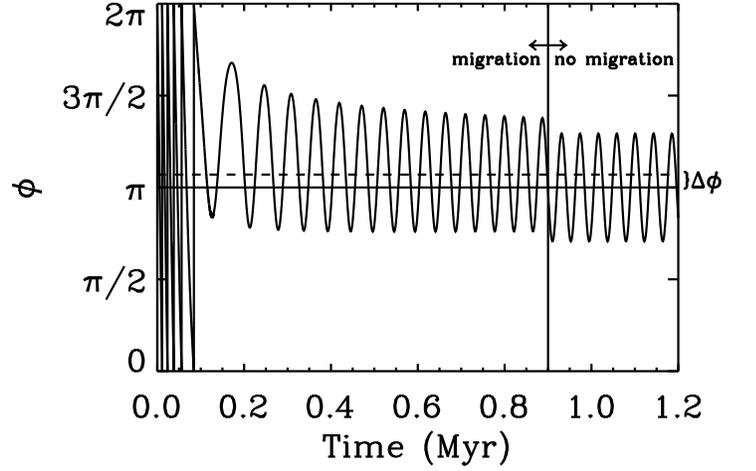}}
\caption{Migration-induced offset in the stable point of symmetric libration. Over the first 0.9 Myr shown, the planet migrates; over the remainder of the plot, the planet's orbit is fixed. The dashed line indicates the offset due to migration, $\Delta\phi$, calculated using Equation \ref{eqn-deltaphifinal} at the plotted time of 0.6 Myr; the agreement with the numerical integration is excellent. For this integration, we drop the term proportional to $\cos(2\phi)$ in $R_{\rm avg,2}$ and set $m\subN$ equal to the mass of Neptune. The planet migrates according to Equation \ref{eqn-migration}, with $t_0 = 10$ Myr and $a_{{\rm N},0} = 23.1$ AU.  Time on the plot is measured from $t=10.3$ Myr when the particle is characterized by $\phi=0$, $e=0.04$, and $a=37.67$ AU (1 AU outside of nominal resonance).}
\label{fig-offset}
\end{figure}

\placefigure{fig-offset}

A key ingredient in understanding asymmetric capture is the fact that for symmetric libration, planetary migration induces an offset, $\Delta\phi$, in the libration center, so that the stable point lies at $\phi=\pi+\Delta\phi$.  Goldreich (1965) finds that to first order in the eccentricity of the particle, symmetric libration is described by the formula
\beq\label{eqn-gold1}
\frac{d^2(\delta\phi)}{dt^2}=-\gamma^2\delta\phi-\frac{dn\subN}{dt} \,\, ,
\eeq
where $\delta\phi = \phi-\pi$ is small, and $2\pi/\gamma$ is the libration period.  The term $-dn\subN/dt$ arises from migration; since $\phi = 2\lambda-\lambda\subN-\pomega$, $\ddot\phi$ due to migration equals $-\ddot\lambda\subN = -\dot n\subN$.  Equation \ref{eqn-gold1} integrates to
\beq\label{eqn-gold2}
\delta\phi = \theta\sin{\gamma t}-\frac{1}{\gamma^2}\frac{dn\subN}{dt} \,\, ,
\eeq
where $\theta$ is the amplitude of libration. Equation \ref{eqn-gold2} describes libration of $\phi$ about $\pi+\Delta\phi$, where the angular offset
\beq\label{eqn-deltaphi1}
\Delta\phi = -\frac{1}{\gamma^2}\frac{dn\subN}{dt}\,\, . 
\eeq
When Neptune migrates outward, $\Delta\phi$ is positive.  The magnitude of the offset is
\beq\label{eqn-deltaphi2}
\Delta\phi = \frac{3}{4\pi}\frac{T_{\rm l}^2}{T_{\rm m}T_{\rm o}} \,\, ,
\eeq
where $T_{\rm l}$ is the libration period, $T_{\rm m} = a\subN/\dot a\subN$ is the migration timescale, and $T_{\rm o}$ is Neptune's orbital period.  
The libration timescale equals
\beq\label{eqn-libtime}
T_{\rm l} \approx C T_{\rm o}\sqrt{\frac{m_\sun}{m\subN}}e^{-1/2} \,\, ,
\eeq
where $C(\alpha) = 1/\sqrt{6\alpha^3[2f_{31}(\alpha)-\alpha^{-2}]} \approx 0.9$.  Combining these expressions, we find that the offset angle is
\beq\label{eqn-deltaphifinal}
\Delta\phi \approx \frac{3C^2}{4\pi}\frac{T_{\rm o}}{T_{\rm m}}\frac{m_\sun}{m\subN}\frac{1}{e} \,\, .
\eeq
In Equations \ref{eqn-libtime} and \ref{eqn-deltaphifinal}, $e$ should be evaluated at exact resonance.


We can verify this offset numerically.  The Hamiltonian given in Equation \ref{eqn-hamavg} with $R_{\rm avg}$ expanded to first order in $e$ reproduces symmetric libration qualitatively correctly (\textsection\ref{sec-expanded}).
We integrate the equations of motion for this Hamiltonian, again forcing Neptune to migrate according to Equation \ref{eqn-migration}.\footnote{For this integration, we drop only the term proportional to $\cos(2\phi)$ from Equation \ref{eqn-expand}; this permits ease of comparison with later integrations in which we restore this term.} By inserting Equation \ref{eqn-migration} into Equation \ref{eqn-deltaphifinal}, we see that our prescription for migration yields an offset, $\Delta\phi$, that is approximately constant with time, modulo the effect of the changing eccentricity of the KBO. Figure \ref{fig-offset} displays a sample time evolution for $\phi$ and verifies that this is the case.  The center of symmetric libration is shifted above $\pi$ by an amount, $\Delta\phi$, that agrees with Equation \ref{eqn-deltaphifinal}. The input parameters for the integration are supplied in the caption to Figure \ref{fig-offset}.

\begin{figure}
\epsscale{1.2}
\plotone{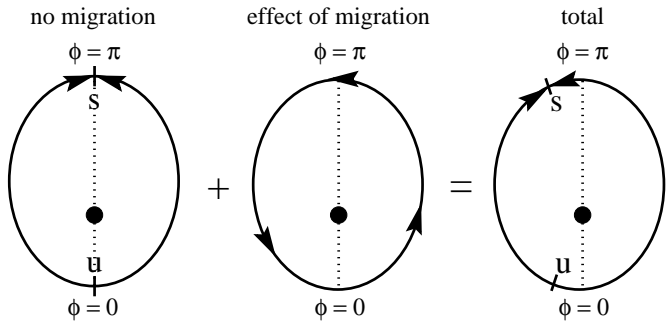}
\caption{Schematic of the accelerations of $\phi$ for symmetric libration.  The letters `s' and `u' mark stable and unstable equilibria, respectively.  When Neptune is not migrating, $\phi$ accelerates toward $\pi$ (left). Neptune's migration accelerates $\phi$ toward larger values (center).  These contributions combine to yield acceleration toward a stable point greater than $\pi$, and acceleration away from an unstable point less than $2\pi$ (right).}
\label{fig-symbalance}
\end{figure}

\placefigure{fig-symbalance}

Physically, the offset occurs because as $a\subN$ increases, the planet's angular speed decreases.  Each conjunction occurs later than it would in the absence of migration; migration accelerates $\phi$ toward larger values.  As a result, the stable point of $\phi$ increases; see Figure \ref{fig-symbalance} for a schematic description.  The offset is akin to the shift in the equilibrium position of a spring in the presence of a constant gravitational field.

\begin{figure}
\epsscale{1.4}
\hbox{\hspace{-0.4in}\plotone{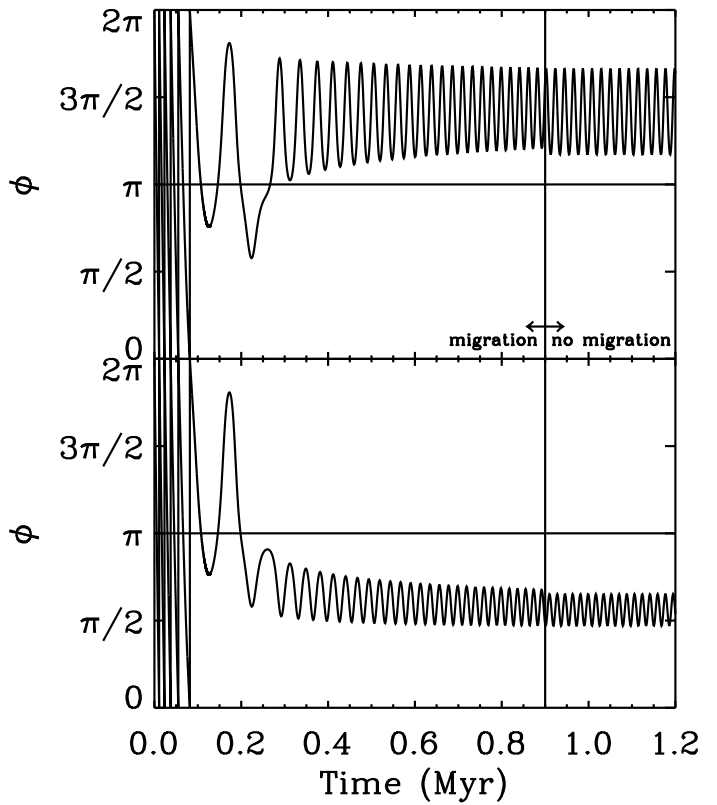}}
\caption{Migration-induced offsets in asymmetric libration.  Two evolutions of $\phi$ exhibiting asymmetric libration are shown, one ending in the trailing island ($\phi > \pi$, top), and the other ending in the leading island ($\phi < \pi$, bottom).  The planet migrates from time 0 to 0.9 Myr and subsequently its orbit is fixed.  Both centers of asymmetric libration are shifted to larger values during the migration.  The migration parameters are the same as those in Figure \ref{fig-offset} except that here,  $R_{\rm avg}$ is expanded to second order in $e$ and at the plotted time of 0, $\phi_0 = 6.19$ (top) and $6.22$ (bottom).}
\label{fig-asymshifts}
\end{figure}

\placefigure{fig-asymshifts}

\begin{figure}
\epsscale{1.2}
\plotone{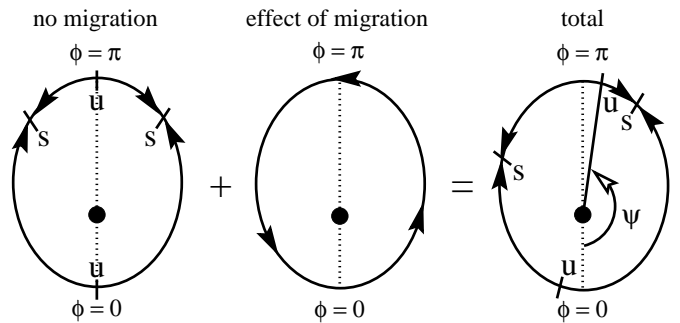}
\caption{Schematic of the accelerations of $\phi$ for asymmetric libration.  The letters `s' and `u' mark stable and unstable equilibria, respectively.  When Neptune is not migrating, $\phi$ accelerates toward the two stable points (left). Neptune's migration accelerates $\phi$ toward larger values (center).  These effects combine to shift the two stable points to larger $\phi$ and the two unstable points to smaller $\phi$ (right).  The angle $\psi<\pi$ marks the location of one of the shifted unstable equilibrium points.  Note how the stable and unstable points are squeezed together at $\phi<\pi$.}
\label{fig-asymbalance}
\end{figure}

The stable points of asymmetric libration are also shifted forward by outward migration.  Figure \ref{fig-asymshifts} displays integrations, using $H_{\rm avg,2}$, for which the migration is turned off after capture into asymmetric resonance.  The libration centers in the absence of migration relax to smaller values of $\phi$.  Figure \ref{fig-asymbalance}, analogous to Figure \ref{fig-symbalance}, explains the shifts pictorially.

\placefigure{fig-asymbalance}

\subsection{Deciding Between the Two Libration Centers}
\label{sec-decide}

Keeping in mind that all libration centers---both for symmetric and asymmetric resonance---shift due to Neptune's migration, we now turn to the question of which island of asymmetric libration captures a greater proportion of particles.  We will refer to the asymmetric libration center with $\phi < \pi$ as the ``leading'' center and to the center with $\phi > \pi$ as the ``trailing'' center. We choose this terminology because in a snapshot of an ensemble of resonant particles (see CJ), particles librating about the leading center ($\phi<\pi$) both attain perihelion and spend a greater fraction of the synodic period at longitudes greater than that of Neptune (see Figure \ref{fig-indirect21}).  As a consequence, these particles appear to cluster at longitudes greater than that of Neptune. Conversely, particles librating about the trailing center ($\phi > \pi$) cluster at longitudes less than that of Neptune.  

We identify three separate factors that determine the relative capture probabilities.

\subsubsection{Migration-Induced Offsets Favor Trailing Island}
\label{sec-asymcap}

If, as a particle transitions from symmetric to asymmetric libration, $\phi$ is greater than some angle $\psi$, then the particle is caught into the trailing island, while if $\phi < \psi$, the particle is caught into the leading island.  At first glance, one might expect that $\psi = \pi$, but in fact $\psi < \pi$.  Figure \ref{fig-asymbalance} illustrates how the addition of the acceleration of $\phi$ due to migration, $\ddot\phi = -\dot n\subN > 0$,  shifts the two stable equilibria to larger $\phi$ and the unstable equilibria to smaller $\phi$.  The unstable equilibrium point that was located at $\phi=\pi$ in the absence of migration is shifted by migration to a value $\psi < \pi$.

To estimate $\psi$ analytically, we perform calculations that are analogous to those that yield Equation \ref{eqn-gold1} and that incorporate the second-order resonant term in the disturbing function.  For small $\delta\phi = \phi - \pi$,
\beq\label{eqn-asymaccel}
\frac{d^2(\delta\phi)}{dt^2}\approx-A\gamma^2\delta\phi-\frac{dn\subN}{dt} \,\, ,
\eeq
where $A(\alpha,e) = 1-8ef_{53}(\alpha)/[2f_{31}(\alpha)-\alpha^{-2}] \approx 1-34e$.  For $A>0$, symmetric libration occurs with frequency $A^{1/2}\gamma$ and asymmetric libration is forbidden. The impossibility of asymmetric libration for $e \lesssim 0.03$ ($A \gtrsim 0$) is not an artifact of our second-order expansion of the Hamiltonian.  It is true even under the exact Hamiltonian, as can be seen by careful examination of level diagrams of the Hamiltonian. 

For $A<0$, asymmetric libration becomes possible;  the unstable equilibrium point, located at $\phi = \pi$ in the absence of migration, is shifted by migration to
\beq\label{eqn-psi}
\psi \approx \pi + A^{-1}\Delta\phi < \pi \,\, ,
\eeq
where $\Delta \phi$ is given by Equation \ref{eqn-deltaphi1}.
The coefficient $A^{-1} \approx -2.8$ for $e=0.04$.  

The relative likelihoods of capture into the two islands depend on the time spent in symmetric resonance librating at $\phi > \psi$ and $\phi < \psi$.  Migration shifts the unstable equilibrium point dividing the two islands backward by $A^{-1}\Delta\phi$.  Migration also shifts the center of symmetric libration forward by $\Delta\phi$.   These two effects compound to render capture into the trailing island more likely.

\begin{figure}
\epsscale{1.4}
\hbox{\hspace{-0.4in}\plotone{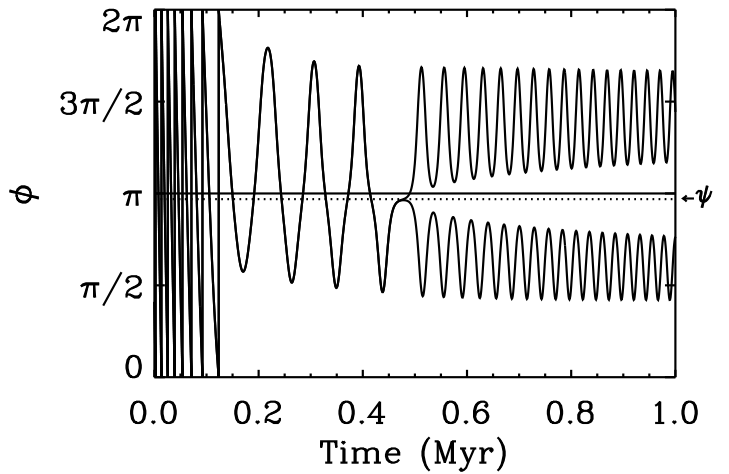}}
\caption{Migration-induced offset in the unstable point dividing the two islands of asymmetric resonance.  Shown are two evolutions of $\phi$ with slightly different initial values of $\phi$. The solid horizontal line marks $\phi=\pi$, and the dotted line marks $\phi=\psi$, evaluated according to Equation \ref{eqn-psi} at the plotted time of 0.48 Myr.  If $\phi>\psi$ at the moment the particle transitions from symmetric to asymmetric resonance, then the particle is caught into the trailing island ($\phi>\pi$); if $\phi<\psi$, then the particle is caught into the leading island ($\phi<\pi$).  The location of the dividing angle inferred from these integrations is well predicted by Equation \ref{eqn-psi}.  Note also how the symmetric librations near the plotted time of $\sim$0.3 Myr are shifted above $\phi=\pi$; this is the same effect documented in Figures \ref{fig-offset} and \ref{fig-symbalance} and \textsection\ref{sec-offset}.  For this integration, $m\subN$ is set equal to the mass of Neptune. The planet migrates according to Equation \ref{eqn-migration}, with $t_0 = 15$ Myr and $a_{{\rm N},0} = 23.1$ AU.  Time on the plot is measured from $t=15.45$ Myr when the particles are characterized by $\phi= 1.2873$ and $1.2876$, $e=0.04$, and $a=37.67$ AU (1 AU outside of nominal resonance).}
\label{fig-transition}
\end{figure}

\placefigure{fig-transition}

We can verify numerically the value of the transition angle, $\psi$.  Figure \ref{fig-transition} displays two evolutions of $\phi$ that start with slightly different initial values for $\phi$.  The solid horizontal line marks $\phi=\pi$, and the dashed line marks $\phi=\psi$, evaluated using Equation \ref{eqn-psi} near the time at which the particle trajectories appear to diverge.  The particle that transitions from symmetric to asymmetric resonance when $\phi<\psi<\pi$ is caught into the leading island, while the particle that makes the transition when $\phi>\psi$ is caught into the trailing island.  To the extent that particles in symmetric libration spend more time with $\phi>\psi$ than with $\phi<\psi$---and note how the symmetric librations are shifted above $\phi=\pi>\psi$---we expect more particles to be captured into the trailing island.  Further quantitative analysis is supplied in \S\ref{sec-quant}.

\subsubsection{Libration Amplitude in Symmetric Resonance and Dependence on Initial Eccentricity}
\label{sec-eccent}

\begin{figure*}
\epsscale{1.2}
\hbox{\hspace{-0.2in}\plotone{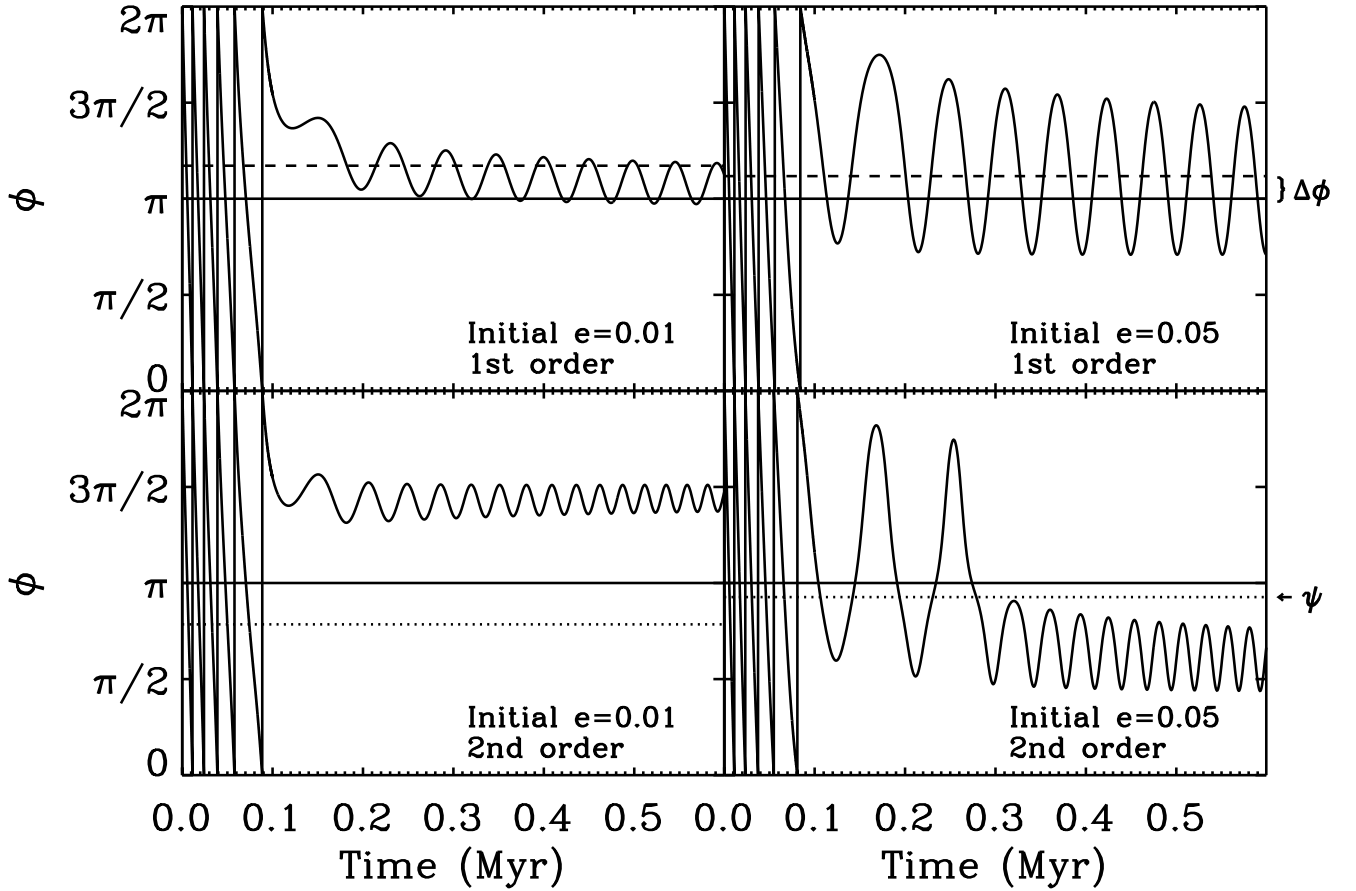}}
\caption{Effect of initial eccentricity on the evolution of $\phi$. Smaller initial eccentricities yield smaller amplitudes of libration and greater likelihoods of capture into the trailing island ($\phi > \pi$). Initial eccentricities equal 0.01 in the left two panels and 0.05 in the right two panels.  In the top two panels, $\phi$ evolves under the potential in which the term proportional to $\cos(2\phi)$ is dropped, yielding symmetric libration.  The dashed lines indicate the value of $\pi+\Delta\phi$ evaluated using Equation \ref{eqn-deltaphifinal} at the plotted time of 0.2 Myr.  In the bottom two panels, the potential is expanded to second order in $e$, yielding the possibility of asymmetric libration.  The dotted lines indicate the value of $\psi$ calculated using Equation \ref{eqn-psi} at a time of 0.2 Myr.  Solid horizontal lines mark $\phi=\pi$.  Remaining initial conditions are the same for all four panels: the planet has the mass of Neptune and migrates according to Equation \ref{eqn-migration}, with $t_0 = 10$ Myr and $a_{{\rm N},0} = 23.1$ AU.  Time on the plots is measured from $t=10.3$ Myr when the particles are at $\phi= 0.126$ and $a=37.67$ AU (1 AU outside of nominal resonance).}
\label{fig-eccent}
\end{figure*}

The amplitude of libration exhibited by a particle in symmetric resonance depends on the particle's eccentricity prior to resonance capture, with larger initial eccentricities producing larger libration amplitudes.  Larger libration amplitudes in symmetric resonance imply that the migration-induced offsets described in \textsection\ref{sec-offset} and \textsection\ref{sec-asymcap} are less effective at changing the ratio of the times spent at $\phi<\psi$ and $\phi>\psi$.  We therefore expect the capture probabilities between the two islands to equalize with larger initial eccentricities.  Figure \ref{fig-eccent} illustrates the effect.  Note that larger eccentricities not only produce larger amplitudes of libration, but also reduce the magnitudes of the offsets $\Delta\phi$ and $\pi-\psi$, as is evident from Equations \ref{eqn-deltaphifinal} and \ref{eqn-psi} (the latter for $e \gtrsim 0.03$).

\placefigure{fig-eccent}

\subsubsection{Uneven Libration Rates and Reversal of Capture Asymmetry}
\label{sec-uneven}

We have argued that the ratio of particles captured into the trailing and leading islands is determined by the ratio of times a particle spends with $\phi>\psi$ and with $\phi<\psi$ while in symmetric libration.  These times are determined not only by the range of values spanned by $\phi$ during symmetric libration, but also by the rate at which the particles librate, i.e., $\dot\phi$ as a function of $\phi$. Immediately before the transition from symmetric to asymmetric libration, a particle librates more slowly when $\phi<\psi$ than when $\phi>\psi$, opposing the tendency to spend more time at $\phi>\psi$ as described in \textsection\ref{sec-asymcap}.  Surprisingly, this effect can be large enough to favor the leading island for some initial conditions and to
reverse the usual sense of the capture asymmetry.

\begin{figure}
\epsscale{1.4}
\vspace{-0.3in}
\hbox{\hspace{-0.35in}\plotone{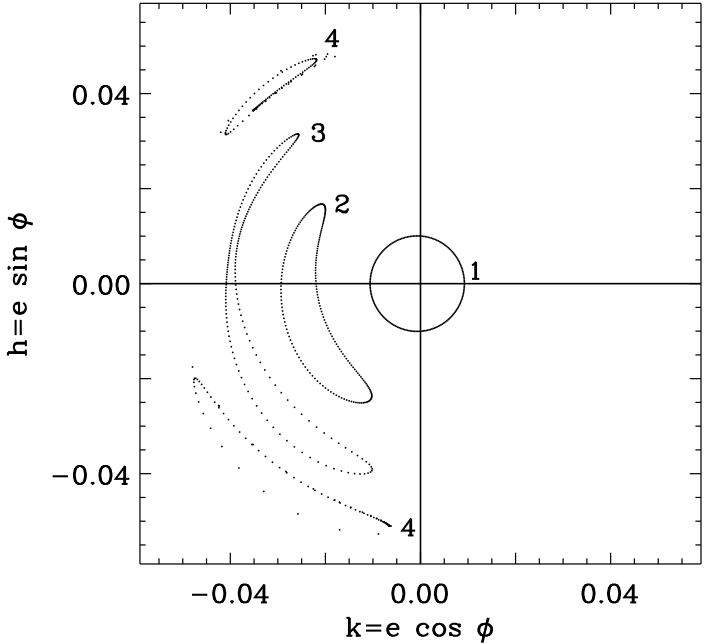}}
\caption{Evolution from circulation (curve 1) to symmetric libration (curves 2 and 3) to asymmetric libration (curve 4) of a collection of 200 particles that initially share the same value of the Hamiltonian. Each point represents
the phase space position of a single particle recorded at 1 of 4 instants.
Neptune migrates according to Equation \ref{eqn-exponential}, with
$\tau = 45$ Myr.
Snapshots are taken at times $t = 2.75$ Myr (curve 1), 4.40 Myr
(curve 2), 4.73 Myr (curve 3), and 5.10 Myr (curve 4).
At $t=0$,  $e\approx 0.01$ and $a\approx 37.67$ AU (1 AU outside of nominal resonance) for each particle. (Since level curves of the Hamiltonian are not exact
circles even in circulation, initial eccentricities and semi-major axes of
particles on a level curve are slightly different.) Note, in curve 3,
the tendency of particles to cluster at $\phi < \pi$,
despite the greater range of values accessible at $\phi > \pi$.}
\label{fig-snapshots}
\end{figure}

Figure \ref{fig-snapshots} provides a series of snapshots of an ensemble of particles beginning in circulation and ending in asymmetric resonance.
For this integration, we use the more realistic migration prescription,
\beq\label{eqn-exponential}
a\subN(t) = a_{\rm N,f} - (a_{\rm N,f} - a_{\rm N,0}) e^{-t/\tau}  \,\, ,
\eeq
where $a_{\rm N,0} = 23.1$ AU is Neptune's initial semi-major axis, $a_{\rm N,f} = 30.1$ AU is Neptune's final semi-major axis, and $\tau$ is a time constant.
In circulation (curve 1),
particles are evenly distributed over an approximate circle,
reflecting the rough constancy of $\dot\phi$.
In symmetric libration (curve 2), $\phi$ changes more slowly near
the turning points of the libration. This slowing is evident as a
clumping of particles near extrema of $\phi$.

\placefigure{fig-snapshots}

Immediately preceding the transition from symmetric to asymmetric resonance (Figure \ref{fig-snapshots}, curve 3), the level curve of symmetric libration deforms to just
surround the islands of asymmetric libration into which the particles will eventually be caught.  The leading island is smaller than the trailing island
due to migration-induced offsets in
the stable and unstable equilibria (see Figure \ref{fig-asymbalance}).
Despite the smaller size of the leading island, however, particles still tend
to clump in its vicinity, reflecting smaller values of $\dot\phi$ there than
near the trailing island. The result, after the last contour of symmetric
libration finally fissions (curve 4), is a tendency to capture more particles
into the leading island.

The greater degree of slowing near the leading island may be crudely
understood by noting that at $\phi < \psi$, the unstable and stable points
are closely juxtaposed (Figure \ref{fig-asymbalance}).  Since both points
correspond to $\ddot\phi = 0$, the potential in this region of phase
space should be relatively flat. Particles having small $\dot\phi$ near this
turning point of symmetric libration
spend a comparatively long time traversing a nearly flat potential.

\subsection{A Sample Quantitative Estimate}
\label{sec-quant}
	
We use our analytic model to estimate
the minimum migration speed above which 
capture into the trailing island is overwhelmingly preferred
over capture into the leading island. Capture into the leading
island is impossible if, at the time of transition from symmetric
to asymmetric resonance, the libration angle $\phi$ fails
to attain values less than $\psi$, i.e., if the sum of both
migration-induced offsets is greater than the amplitude
of symmetric libration:

\begin{equation}
\Delta \phi + (\pi - \psi) \gtrsim \theta \, .
\label{inequality}
\end{equation}

\ni All quantities in this expression should be
evaluated at the time of capture into asymmetric resonance.
Using Equations \ref{eqn-deltaphifinal} and \ref{eqn-psi}, we re-write the above criterion as 

\begin{equation}
T_{\rm m} \lesssim \left(1 - A^{-1} \right) \frac{3C^2}{4\pi} \frac{m_{\odot}}{m_{\rm N}} \frac{T_{\rm o}}{e\theta} \, .
\label{inequality2}
\end{equation}

\ni We may estimate $\theta$ as a function of $e$ in the adiabatic limit. In this limit, at the time of transition into asymmetric libration, the contour of constant Hamiltonian on which the particle lies (the ``asymmetric separatrix'') contains an unstable point at $\phi=\pi$.  On the asymmetric separatrix, the turning points of libration have approximately the same value of $e$ as the value at $\phi=\pi$.  Then Equation \ref{eqn-hamilt} requires that
\begin{multline}
\alpha(f_{31}e\cos\phi_{\rm tp} + f_{53}e^2\cos 2\phi_{\rm tp})-\frac{1}{2\alpha}e\cos\phi_{\rm tp} \approx \\
\alpha(f_{31}e\cos\pi + f_{53}e^2\cos 2\pi)-\frac{1}{2\alpha}e\cos\pi
\end{multline}
where $\phi_{\rm tp}$ gives the values of $\phi$ at the turning points:
\beq\label{eqn-phitp}
\phi_{\rm tp} = \cos^{-1} \left[1-\frac{B(\alpha)}{e}\right]
\eeq
and $B(\alpha)\equiv [\alpha f_{31}-1/(2\alpha)]/[2\alpha f_{53}] \approx 0.058$.  The amplitude of libration is
\beq\label{eqn-amp}
\theta(e) = |\pi-\phi_{\rm tp}(e)| \,\, .
\eeq

For $m_{\rm N}/m_{\odot} = 5 \times 10^{-5}$ and $a_0 =37.67$ AU, 
the integration displayed in Figure \ref{fig-snapshots} offers the value $e \approx 0.04$ at the time of transition from symmetric to asymmetric resonance.  Equation \ref{eqn-amp} then yields $\theta \approx 1.1$, consistent with the amplitude shown in Figure \ref{fig-snapshots}.
From these values, we calculate that the critical value
of $T_{\rm m}$ below which capture into
the trailing island is overwhelmingly preferred is
$T_{\rm m, crit} \sim 3.8 \times 10^7 \yr$. In the next
section on numerical integrations, we check the accuracy
of this estimate.

The only remaining parameter for which we have yet to supply an analytic expression is the eccentricity at the transition from symmetric to asymmetric libration.  To estimate this, one could work again in the adiabatic limit, using the adiabatic invariant, $J \equiv \oint\Gamma d\phi$ (approximately proportional to the area enclosed by the path of a particle in $h$-$k$ space in the small $e$ limit).  The value of $J$ in symmetric libration equals the value of $J$ appropriate to the asymmetric separatrix that the particle crosses. One could exploit this fact to identify that separatrix and its associated value of $e$.
Another improvement would be to compute
the relative times that $\phi$ spends above and below $\psi$
when Equation \ref{inequality} is not satisfied.
Such a calculation would have to account for the anharmonicity 
of motion in asymmetric libration (see, e.g., Malhotra 1996),
an anharmonicity that is only accentuated when the perturber
migrates (see \S\ref{sec-uneven}).

	We do not pursue these additional
computations analytically, but rather content ourselves with
our physical, order-of-magnitude understanding of
asymmetric capture embodied in Equations \ref{inequality2}--\ref{eqn-amp}
and proceed with more numerical explorations in the next section.
We note in passing that the resonance capture theory of Henrard (1982)
and the analytic estimates of capture probabilities derived
therefrom (Borderies \& Goldreich 1984)
cannot be applied to the problem of asymmetric capture
without substantial modification. This is because such theories
are written in the strict adiabatic
limit in which the migration timescale is infinitely
longer than the libration timescale. But asymmetric
capture relies upon a finite migration timescale.
In other words, migration-induced offsets are absent
from the theory of Henrard (1982).

\subsection{Population Ratios as a Function of Migration Speed and Initial Eccentricity}
\label{sec-ratios}

\begin{figure}
\epsscale{1.4}
\hbox{\hspace{-0.4in}\plotone{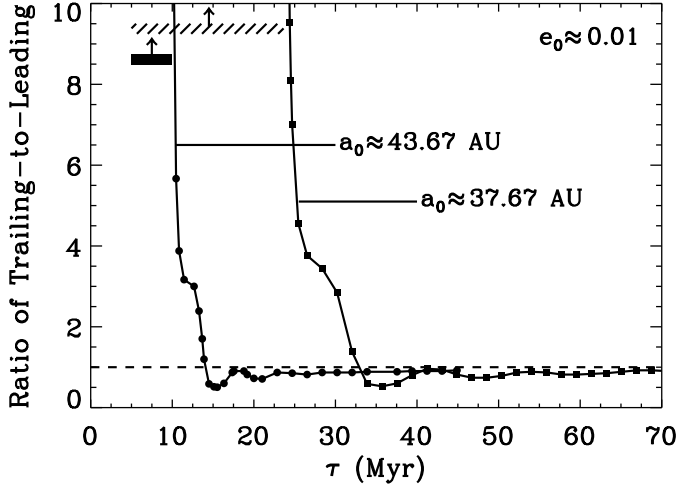}}
\caption{Dependence on the exponential migration timescale, $\tau$, of the ratio of particles captured into trailing vs.~leading asymmetric resonance assuming low initial eccentricities ($e_0 \approx 0.01$).  Horizontal (solid and hatched) bars correspond to a ratio of $\infty$.  For fast migration, more particles are caught into the trailing island, and for slow migration, the ratio is near 1.  For all timescales with a ratio plotted, 100\% of particles are captured from circulation.  For this plot, the mass of the planet is Neptune's true mass,  and Neptune migrates according to Equation \ref{eqn-exponential}, with $a_{\rm N,0} = 23.1$ AU and $a_{\rm N,f} = 30.1$ AU. Each ratio is calculated by following 200 particles that initially circulate, have the same Hamiltonian value, and are spaced evenly in $\phi_0$. The hatched bar corresponds to particles having initial semi-major axes of $a_0 \approx 37.67$ AU ($1$ AU outside of nominal resonance), and the solid bar corresponds to $a_0 \approx 43.67$ AU ($7$ AU outside of nominal resonance). Because Neptune's migration rate decreases exponentially with time, Neptune is migrating more slowly when it captures particles that start further from resonance.  As a result, particles having $a_0 \approx 37.67$ AU are asymmetrically captured over a greater range of migration timescales as compared to those having $a_0 \approx 43.67$ AU.}
\label{fig-smalle}
\end{figure}

\placefigure{fig-smalle}

\begin{figure}
\epsscale{1.4}
\hbox{\hspace{-0.4in}\plotone{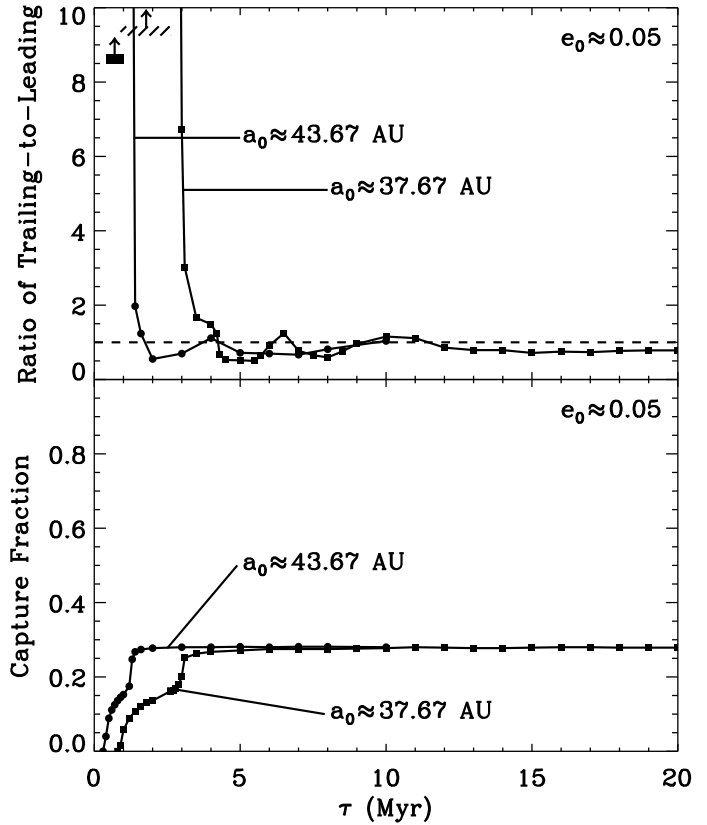}}
\caption{(Top) Dependence on the exponential migration timescale, $\tau$, of the ratio of particles captured into trailing vs.~leading asymmetric resonance assuming high initial eccentricities ($e_0 \approx 0.05$).  Horizontal bars correspond to a ratio of $\infty$. 
Fast migration yields large asymmetries favoring the trailing island, while slow migration yields ratios near 1. The timescale dividing these behaviors is shorter for $e_0\approx0.05$ than for $e_0\approx0.01$ (Figure \ref{fig-smalle}); see \textsection\ref{sec-eccent} for an explanation.  (Bottom) Fraction of particles caught from circulation.  Note that large asymmetric capture ratios correspond to timescales shorter than those producing adiabatic capture fractions.
Migration parameters are the same as those for Figure \ref{fig-smalle}.  Each datum is calculated by following 800 particles that initially circulate, have the same Hamiltonian value, and are spaced evenly in $\phi_0$. For the hatched bar,  $a_0 \approx 37.67$ AU, and for the solid bar, $a_0 \approx 43.67$ AU. Neptune is migrating more slowly when it captures particles that start further from resonance.}
\label{fig-largee}
\end{figure}

\placefigure{fig-largee}

We calculate numerically the ratio of captures into the trailing and
leading islands as a function of migration timescale, initial semi-major axis,
and initial eccentricity,
using the migration prescription given by Equation \ref{eqn-exponential}.
For a given $\tau$, a set of $\eta$ initial values of
($a_0$, $e_0$, $\phi_0$)
are computed which correspond to the same values of $H_{\rm avg,2}$
and $N$. If particles do not all start on the same Hamiltonian
level curve, phase differences accumulate between particles and complicate interpretation of the final capture
ratio. Initial values of $a_0$ and $e_0$ are chosen such that
all particles begin in circulation; initial values of $\phi_0$ are
distributed uniformly from $0$ to $2\pi$. To achieve sufficient
statistics, we select $\eta = 200$ or $800$ depending on the efficiency of capture from circulation. Figure \ref{fig-smalle}
plots capture ratios for particles with small initial eccentricity
($e_0 \approx 0.01$), while Figure \ref{fig-largee} (top) supplies
results for larger initial eccentricity ($e_0 \approx 0.05$).
In each figure, two curves for two different values of $a_0$
are delineated, one starting approximately 1 AU away from resonance, and the other
starting 7 AU away.

All three effects documented in \textsection\ref{sec-decide} manifest
themselves in Figures \ref{fig-smalle} and \ref{fig-largee} (top). Shorter migration
timescales yield greater asymmetries in the capture ratios, with
preference given to capture into the trailing island, because
of migration-induced offsets in the stable and unstable points
of symmetric and asymmetric resonance (\textsection\ref{sec-asymcap}).
For $e_0 \approx 0.01$, $\tau \lesssim 10$ Myr, and our chosen $a_0$'s, the
probability of capture into the trailing island is overwhelming.
Particles that originate from smaller semi-major axes are more likely
to exhibit asymmetric capture because they are caught at earlier
times when Neptune is assumed to have migrated more quickly.
At large $\tau$, the capture ratio approaches unity from below,
a reflection of the nearly flat potential at $\phi < \psi$
(\textsection\ref{sec-uneven}). The ratio of trailing-to-leading
particles never dips below 0.5.

Our expectation, based on a simple calculation in
\S\ref{sec-quant}, that the capture ratio is infinite
if $T_{\rm m} \equiv a_{\rm N}/\dot{a}_{\rm N} \lesssim 38$ Myr
is largely borne out in Figure \ref{fig-smalle}. The relevant curve is the
one for $a_0 \approx 37.67$ AU, for which parameter values best match
those assumed for our calculation. The capture ratio is infinite
for $\tau \lesssim 23$ Myr, which translates to
$T_{\rm m} = \tau a_{\rm N}(t)/[a_{\rm N,f}-a_{\rm N}(t)] < 87$ Myr,
a critical timescale that differs by a factor of 2.3 from our
estimate in \S\ref{sec-quant}. The discrepancy could arise from our use of Equation \ref{eqn-amp}, which is only correct in the adiabatic limit and for eccentricities that are not small, and our use of Equation \ref{eqn-libtime}, which is only correct in the limit of small libration amplitude.

As initial eccentricities of particles increase,
capture into symmetric resonance becomes no longer certain.\footnote{Capture is certain if the separatrix dividing circulation and symmetric libration forms outside of the circulation trajectory occupied by the particle and probabilistic if this separatrix forms inside the circulation trajectory (see, e.g., Peale 1986).} For $e_0 \approx 0.05$, the fraction of captured objects saturates at 28\% for $\tau > 5$ Myr and decreases with  shorter $\tau$ (Figure \ref{fig-largee}, bottom).  Of those that
are captured, strong asymmetries in the trailing-to-leading ratios do not appear except at $\tau\sim1$ Myr because
the larger amplitudes of symmetric libration caused by the larger
initial (free) eccentricities weaken the effects of migration-induced shifts (\textsection\ref{sec-eccent}). Moreover, the shifts themselves are smaller because of the larger eccentricities.  For example, for $\tau = 10$ Myr and $e_0\approx0.01$, the ratio of trailing-to-leading particles can be infinite, while for the same $\tau$ and $e_0\approx0.05$, the ratio is near unity.  The curves for $e_0\approx0.05$ are shifted to lower $\tau$'s compared to those for $e_0\approx0.01$ by a factor of $\sim$8.

The computations we have just described and numerical simulations by CJ agree that the characteristic migration timescales required to produce strong asymmetric capture are in the range $10^6$--$10^7$ yr.  Equation \ref{inequality2} also suggests the same critical timescales.  Our calculations yield larger asymmetries, however, than the purely numerical ones of CJ.  For example, Figures \ref{fig-smalle} and \ref{fig-largee} indicate that at $\tau = 10^6$ yr, the capture ratio is infinite, while CJ report a ratio of 3-to-1. We are unsure as to why there are differences, but the computations are at heart different: one employs a truncated version of the Hamiltonian and time-averages over the synodic period, while the other has no such approximations.  While the semi-analytic model has served us well in illuminating the numerous physical effects underlying what was once a purely numerical result, we should probably rely on the numerical simulations for actual comparisons between theory and observation.

\section{COMPARISON WITH OBSERVATIONS}
\label{sec-observations}
\begin{figure*} 
\vspace{-1in}                                                               
\epsscale{1.0}
\hbox{\hspace{+0.15in}\plotone{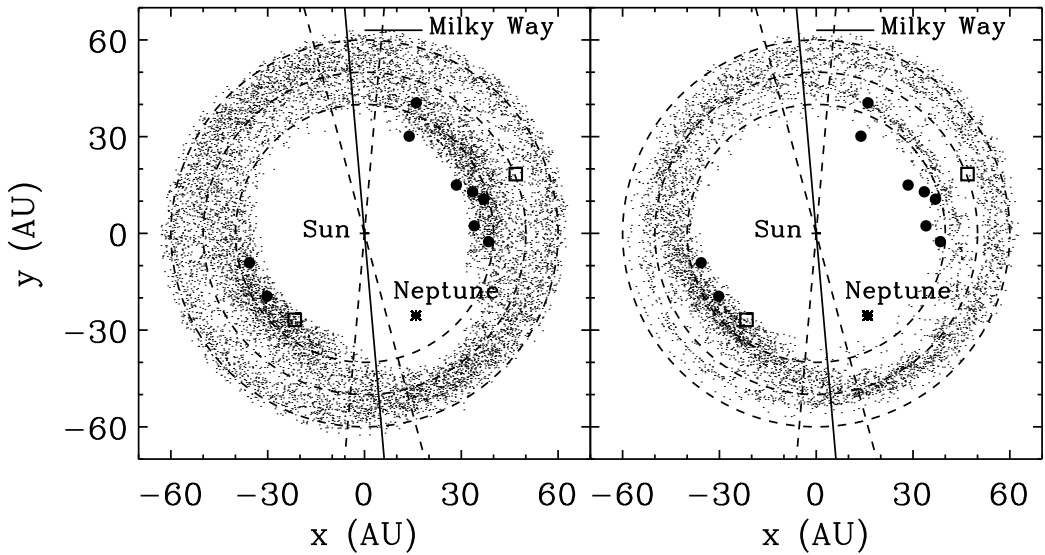}}
\caption{Current positions of 11 known Twotinos discovered by surveys world-wide. Of these 11, 2 are symmetric librators (open squares), while 9 are asymmetric librators (solid circles) that we use to constrain Neptune's ancient migration speed. Overlaid are simulated snapshots of 2:1 resonant objects taken from CJ; the left panel portrays the outcome for an exponential migration timescale of $\tau=10^7$ yr, while the right panel corresponds to $\tau=10^6$ yr. Two asymmetric librators lie in the trailing island, while seven lie in the leading island; a de-biased estimate of the population ratio is 3-to-6 and rules out the hypothesis that $\tau\leq10^6$ yr at 99.65\% confidence.}
\label{fig-double}
\end{figure*}

Do the observations indicate symmetry or asymmetry in the distribution of 2:1 resonant KBOs with respect to the Sun-Neptune line? The Deep Ecliptic Survey (Chiang et al.~2003ab; Buie et al.~2003; Elliot et al.~2004) has developed a classification scheme that identifies resonant KBOs by virtue of their librating angles; an object is deemed resonant if three orbital integrations, each lasting 30 Myr and starting with initial conditions lying within the 3$\sigma$ confidence surface of possible osculating orbits, all yield libration of the same resonance angle. By this criterion, 11 2:1 resonant KBOs (``Twotinos'') are identified, of which 2 librate symmetrically and 9 librate asymmetrically. Of the asymmetric librators,   
2 librate about $\phi > \pi$ (trailing island), and               
7 librate about $\phi < \pi$ (leading island).                                The current positions of all 11 Twotinos are displayed in Figure \ref{fig-double}; they are overlaid on theoretical snapshots of Twotinos taken from CJ.                                 

\placefigure{fig-double}             
                                                                              
The observed asymmetry---2-to-7, in favor of the leading island---argues against a rapid migration history for Neptune. Given the initial conditions assumed by CJ---most notably initial eccentricities ranging uniformly from 0 to 0.05---the hypothesis that Neptune's migration timescale was as short as $a_{\rm N}/\dot{a}_{\rm N} \sim 10^6$ yr predicts that the probability that a particle caught into asymmetric resonance is in the trailing island is $u\approx0.75$ (CJ), with shorter migration timescales presumably giving rise to greater values of $u$.  Given the observed (possibly biased---see below) asymmetry of 2-to-7, the probability that $u\geq0.75$ ($\tau \leq 10^6$ yr) is 0.04\%, where we have used the differential probability distribution $dP/du = (S+1)S!/[Y!(S-Y)!]u^Y(1-u)^{S-Y}$ (Port 1994, pages 264--265).\footnote{Chapter 23 on random Bernoulli trials contains typos in the formulae for $dP/du$.}  Here $S=9$ is the sample size and $Y=2$ is the number of observed trailing objects.  Thus, $\tau\leq 10^6$ yr seems unlikely.  The probability that $u\geq 0.5$ ($\tau \leq 10^7$) is 5.5\%.   

\subsection{Theoretical Caveats}
One theoretical caveat to the above interpretation is that the capture probabilities into the two islands are sensitive to the particles' initial eccentricities (see \textsection\ref{sec-eccent} and \textsection\ref{sec-ratios}).  The distribution of eccentricities assumed by CJ might not be realistic; see, e.g., Chiang et al.~(2003ab) who present evidence that a fraction of the Kuiper belt was pre-heated to large eccentricities prior to resonance sweeping.  If Twotinos today originated from larger eccentricity orbits prior to resonance capture, then migration timescales as short as $\sim$$10^6$ yr would yield capture ratios closer to unity.  However, objects on initially larger eccentricity orbits are less likely to be caught at all by the 2:1 resonance.

If we assume that Twotinos originated from circulating orbits having initial eccentricities $\lesssim 0.05$, then slow migration, on timescales longer than $\sim$$10^7$ yr, is a more viable conclusion.  Based on our computations, slow migration yields $0.33\leq u\leq0.5$. Given the observed asymmetry of 2-to-7, the probability that $0.33\leq u\leq0.5$ ($\tau\geq10^7$ yr) is 24\%.

A second theoretical caveat is the possibility
that objects originally caught into one island transition to the other
over the age of the solar system (Hahn, personal communication).
We expect such mixing to afflict only objects having the largest
libration amplitudes. Mixing has been reported by CJ on timescales
as short as $10^7$ yr for objects having libration amplitudes
of 45$^{\circ}$ (see their figure 8). The Twotinos reported
by the Deep Ecliptic Survey do not exhibit mixing on timescales
as long as 30 Myr, but longer-term integrations have yet to be
performed. A third caveat is that massive
Twotinos---objects having a few times Pluto's mass---can
deplete the two islands differentially
(Tiscareno \& Malhotra 2003; Tiscareno 2004). Such large
objects either have yet to be discovered or were
ejected early in the history of the solar system.

\subsection{Observational Biases}
What about observational biases? Since the 3:2 resonance admits no asymmetric libration, Plutinos, unlike Twotinos, have no choice but to be distributed symmetrically about the Sun-Neptune line.  For this reason, Plutinos can serve to calibrate detection efficiencies of surveys as a function of orbital longitude. Observationally, 35 Plutinos have been discovered at longitudes leading Neptune and 25 have been discovered at trailing longitudes (see, e.g., figure 14 of CJ).  The observed asymmetry is in the same sense as for the Twotinos. 
If we assume that Plutinos are equally likely to be trailing or leading Neptune, then the probability that the difference in the number of Plutinos found trailing and leading Neptune is $\geq 10$ (given a sample size of 60) is 25\% according to the binomial distribution.
This result does not indicate strongly whether discovering and astrometrically recovering KBOs at longitudes leading Neptune has historically been easier than at trailing longitudes.\footnote{Leading longitudes correspond to spring/summer in the Northern hemisphere, and trailing longitudes correspond to fall/winter.} If we assume a bias and use the Plutinos to de-bias the Twotinos, then we estimate a de-biased trailing-to-leading ratio for Twotinos of 3-to-6 (2.6-to-6.4, rounded to the nearest integer).  The likelihoods cited above increase to 0.35\% ($\tau\leq 10^6$ yr), 17\% ($\tau\leq 10^7$ yr), and 39\% ($\tau\geq10^7$ yr). A small increase in the total sample size of 2:1 KBOs could dramatically increase our confidence that fast migration did not occur.  We look forward to the advent of the Pan-STARRS synoptic survey with its anticipated discovery of $\sim$$10^3$ Twotinos.


\section{SUMMARY AND EXTENSIONS}
\label{sec-summary}

Orbital migration of bodies embedded in disks can explain observed structures in systems ranging from planetary rings to extra-solar planets. As a body migrates, it can capture others into mean-motion resonances, leaving a potentially lasting signature of its migration.  In our solar system, numerous Kuiper belt objects (KBOs) exist in mean-motion resonance with Neptune, suggesting that Neptune may have migrated outwards by several AUs (Malhotra 1995; Chiang et al.~2003ab).  Chiang \& Jordan (2002, CJ) discover by numerical simulation that the subset of KBOs inhabiting the 2:1 resonance furnishes a unique probe of Neptune's migration history.  The resonant angle, $\phi$, of such KBOs can librate about values equal to $\pi$ (symmetric resonance), greater than $\pi$ (trailing asymmetric resonance), or less than $\pi$ (leading asymmetric resonance). These authors find that if Neptune's migration occurred on timescales shorter than $10^7$ years (for their assumed initial conditions), more 2:1 resonant KBOs would have been captured into trailing resonance than into leading resonance; as a consequence, more KBOs would be discovered today at longitudes trailing Neptune than at longitudes leading it.  For longer timescales of migration, more equal numbers of KBOs would be found leading and trailing Neptune.  Observational confirmation of more trailing than leading KBOs in 2:1 resonance would constitute strong evidence in favor of the (fast) migration hypothesis; while current tallies are more consistent with equal populations, a definitive census should be possible with the upcoming Pan-STARRS survey.

We have explained the physical origin of this capture asymmetry within the context of the circular, restricted, three-body problem.  Asymmetric libration in the 2:1 resonance occurs as a result of the superposition of the direct and indirect components of the planet's perturbing acceleration (see also Frangakis 1973; Pan \& Sari 2004).  The direct component, produced by the direct gravitational attraction between Neptune and the KBO, serves to accelerate the longitude of conjunction ($\phi$) toward $\pi$.  The indirect component arises from changes in the acceleration of the KBO by the Sun brought about by the latter's reflex motion induced by Neptune. This acceleration, whose effect on the KBO's orbit depends only on the difference between the true longitudes of Neptune and of the KBO, accelerates $\phi$ toward $0$ over a synodic period.  Thus, the indirect and direct perturbations can counter-balance each other to produce asymmetric libration of $\phi$ about angles intermediate between $0$ and $\pi$, and intermediate between $\pi$ and $2\pi$.  

When employing a literal series expansion of the disturbing potential, the inclusion of an inappropriate number of terms can cause spurious asymmetric resonances to appear.  For the 2:1 resonance, asymmetric libration is represented qualitatively well by a literal expansion to second order in $e$ but not by expansions to first, third, or fourth order.  The second-order expansion should be regarded as nothing more than a useful fitting formula for the true potential.

Particles caught by an outwardly migrating 2:1 resonance fall first into symmetric libration before evolving to asymmetric libration.  At the time of transition from symmetric to asymmetric resonance, if the value of $\phi$ for a particle is larger than a critical angle $\psi$, then the particle will be caught into the trailing island of asymmetric libration, while if $\phi<\psi$, it will be caught into the leading island.  The relative probability that a particle is captured into the trailing rather than the leading island is determined by the fraction of its time spent during symmetric libration at $\phi$'s greater than and less than $\psi$.

Three factors determine this fractional time spent:
\begin{enumerate}
\item Migration-induced shifts of the stable and unstable equilibrium points of the resonant potential.
Planetary migration shifts the stable points for symmetric and asymmetric resonance to larger values.  For asymmetric resonance, migration shifts the unstable point formerly at $\pi$, to a smaller value, $\psi<\pi$.  Analytic theory informs us that the angular offsets of the stable point of symmetric libration and of the unstable point of asymmetric libration vary inversely with migration timescale, $T_{\rm m} = a\subN/\dot a\subN$. 
The greater the shifts, the more likely it is that particles spend more time in symmetric resonance at $\phi>\psi$ and are captured into the trailing island.  If the sum of both shifts exceeds the amplitude of symmetric libration---true if $T_{\rm m} \lesssim T_{\rm m, crit} \sim 10^7$ yr---preference for capture into the trailing island is overwhelming.
\item Initial particle eccentricity.
The influence of migration-induced shifts is greatest when the amplitude of symmetric libration is smallest; the latter demands small particle eccentricities in circulation prior to capture. Analytic theory also tells us that the magnitudes of the shifts themselves grow with smaller eccentricities. 
\item Uneven libration rates during symmetric libration.  
The relative time spent is also affected by differences in $\dot\phi$ over one period of symmetric libration.  The average libration rate at $\phi < \psi$ is lower than at $\phi > \psi$ because the close juxtaposition of two equilibrium points at $\phi < \psi$ flattens the potential there. The difficulty with which a particle traverses the turning point at $\phi < \psi$ counteracts the effect of migration-induced shifts and can even reverse the usual sense of the capture asymmetry for large $T_{\rm m}$, but not in a way that gives the leading island more than a 2-to-1 advantage over the trailing island in attracting occupants.
\end{enumerate}

We confirm the results of CJ that the migration timescales necessary to capture more particles into the trailing than the leading island are between about $10^6$ and $10^7$ yr. For $T_{\rm m}$ shorter than $10^6$ yr, the resonance becomes increasingly unable to capture particles at all.  For a realistic migration prescription (e.g., an exponential prescription such as Equation \ref{eqn-exponential}), Neptune's migration rate decreases with time, so particles beginning at smaller semi-major axes are captured while Neptune is migrating more quickly and are more likely to exhibit asymmetric capture.  The final asymmetry in the longitudes of the particles is sensitive to their initial semi-major axes and, more significantly, their initial eccentricities.

Asymmetric libration is not unique to the 2:1 resonance (e.g., Message 1958; Frangakis 1973; Beaug\'e 1994; Malhotra 1996; Winter \& Murray 1997; Pan \& Sari 2004).  In a $p$:1 exterior resonance, the resonant angle,
\beq
\phi = p\lambda - \lambda\subN - (p-1)\pomega \,\, ,
\eeq
can librate asymmetrically as well.  For all other resonances, including interior resonances (e.g., 1:2), asymmetric resonance does not occur.  Asymmetric libration of the type exhibited by the 2:1 resonance requires a non-zero indirect acceleration.  We prove below that the indirect acceleration always averages to zero over one synodic period for a $p$:$q$ resonance except when $q=1$, where $p>q$ refers to an exterior resonance and $p<q$ to an interior resonance ($p\neq q$ are relatively prime, positive integers).  The order of the resonance is $s=|p-q|$, and the resonant angle is $\phi = p\lambda -q\lambda\subN - (p-q)\pomega$. Our proof is essentially identical to that of Frangakis (1973), and derived independently; we offer our expanded version for ease of reference.

The azimuthal component of the indirect acceleration of the KBO by Neptune equals $-(Gm\subN/a\subN^2)\sin\Delta\theta$, where 
$\Delta\theta \equiv \lambda\subN - f - \pomega$
is the angle between the true longitudes of the planet and of the particle.  The proof for the radial component reads completely analogously.  Over one synodic period, the tangential acceleration time-integrates to
\begin{align}\label{eqn-torqueintegral}
\langle T\rangle &= -\frac{Gm\subN}{a\subN^2} \oint \sin(\lambda\subN-f-\pomega) dt \nonumber \\
& = -\frac{Gm\subN}{a\subN^2} \int_0^{2\pi s} \frac{\sin(\Delta\theta)}{n\subN-\dot f(\Delta\theta)-\dot\pomega} d(\Delta\theta) \,\, .
\end{align}
After one full period of the particle, the particle returns to the same true longitude while the planet increases its true longitude by $2\pi s/q$, increasing $\Delta\theta$ by $2\pi s/q$.  It follows that $\dot f(\Delta\theta)$ is periodic in $\Delta\theta$ with period $2\pi s/q$, as is any function of $\dot f$, including
\beq
F(\Delta\theta) \equiv \left[n\subN-\dot f(\Delta\theta)-\dot\pomega\right]^{-1} \,\, ,
\eeq
which may be Fourier-decomposed as
\beq
F(\Delta\theta) = \frac{b_0}{2} + \sum_{j=1}^\infty b_j\cos\left(\frac{j\pi}{L}\Delta\theta\right) +  \sum_{j=1}^\infty c_j\sin\left(\frac{j\pi}{L}\Delta\theta\right) \,\, ,
\eeq
where $2L \equiv 2\pi s/q$. We have neglected the tiny variation of $\dot\pomega \ll n\subN$ over a synodic period. Plugging this expression for $F$ into Equation \ref{eqn-torqueintegral}, we find
\begin{align}
\langle T\rangle &= -\frac{Gm\subN}{a\subN^2}\int_0^{2\pi s}F(\Delta\theta)\sin(\Delta\theta)d(\Delta\theta) \nonumber \\
&= -\frac{Gm\subN}{a\subN^2}\left[ s\sum_{j=1}^\infty b_j \int_0^{2\pi} \cos(jqx)\sin(sx)dx + \right. \nonumber \\
&\qquad\qquad\quad \left. s\sum_{j=1}^\infty c_j \int_0^{2\pi} \sin(jqx)\sin(sx)dx \right] \nonumber \\
&= -\frac{Gm\subN}{a\subN^2}s\pi\sum_{j=1}^\infty c_j\delta_{(jq)s}  \,\, .
\end{align}
Thus, $\langle T\rangle\neq 0$ only when $jq=s=|p-q|$ for some $j$.  This is satisfied only when $q=1$ and $j=|p-q|$ under our assumption that $p$ and $q$ are relatively prime.\footnote{When $q$ is even, one can see that the indirect term integrates to zero by appealing to diagrams analogous to Figure \ref{fig-indirect21} and noting symmetry over the synodic period.} We conclude that asymmetric libration, of the kind exhibited by the 2:1 resonance, can only exist for exterior $p$:1 resonances.\footnote{The only resonance for which our proof does not apply is the 1:1 resonance. But it is clear that the indirect acceleration does not time-average to zero in tadpole and horseshoe orbits, which may be regarded as asymmetric and symmetric librations, respectively. See Frangakis (1973) and Pan \& Sari (2004).}

Future theoretical work should focus on determining the degree of stochasticity in migration driven by scattering planetesimals, and on the size spectrum of planetesimals required for resonance capture to proceed smoothly. We have made preliminary estimates which suggest that for a Neptune-mass disk in the vicinity of Neptune, the disk mass must be concentrated in planetesimals whose sizes do not exceed $\sim$100 km for the theory that we have described in this paper to apply. Reasons to imagine that planetesimal sizes were small can be found in the review of planet formation by Goldreich et al.~(2004).

\acknowledgements
We thank Re'em Sari and Margaret Pan for pointing out the
importance of the indirect term early in our investigation
of asymmetric resonance. 
Renu Malhotra provided a referee's report that helped to improve the presentation of this paper and that spurred us to greater insights.
We are grateful to Joe Hahn, David Jewitt,
Alessandro Morbidelli, and Mark Wyatt for stimulating discussions.
R.A.M. acknowledges support
by a National Science Foundation (NSF) Graduate Fellowship.
E.I.C. acknowledges support by NSF
Grant AST 02-05892 and the Alfred P.~Sloan Foundation.


\begin{references}
Beaug\'{e}, C.~1994, Celestial Mechanics and Dynamical Astronomy, 60, 225\\
Borderies, N., \& Goldreich, P. 1984, Celest. Mech. 32, 127 \\
Buie, M.W., et al.~2003, Earth, Moon, and Planets, 92, 113\\	
Chiang, E.I., \& Jordan, A.B. 2002, \aj, 124, 3430 (CJ)\\
Chiang, E.I., et al.~2003a, \aj, 126, 430\\
Chiang, E.I., et al.~2003b, Earth, Moon, and Planets, 92, 49\\
Elliot, J.L., et al.~2004, AJ, submitted\\
Fernandez, J.A., \& Ip, W.H.~1984, Icarus, 58, 109\\
Frangakis, C.N. 1973, ApSS, 22, 421\\
Goldreich, P.~1965, \mnras, 130, 159\\
Goldreich, P., Lithwick, Y., \& Sari, R.~2004, \araa, 42, 549\\
Hahn, J.M., \& Malhotra, R.~1999, \aj, 117, 3041\\
Henrard, J. 1982, Celest. Mech., 27, 3 \\
Lee, M.H., \& Peale, S.J.~2002, \apj, 567, 596\\
Malhotra, R.~1995, \aj, 110, 420\\
Malhotra, R.~1996, \aj, 111, 504 \\
Marcy, G., et al.~2001, \apj, 555, 418\\
Message, P.J.~1958, AJ, 63, 443\\
Murray, C.D., \& Dermott, S.F.~1999, Solar System Dynamics (Cambridge: Cambridge University Press)\\
Murray, N., et al.~1998, Science, 279, 69\\
Pan, M., \& Sari, R.~2004, AJ, 128, 1418\\
Peale, S.J. 1986, in Satellites, eds. J.A. Burns \& M.S. Matthews (Tucson: Univ. Arizona Press), 159\\
Port, S.C.~1994, Theoretical Probability for Applications (New York: John Wiley \& Sons, Inc.)\\
Press, W.H., Teukolsky, S.A., Vetterling, W.T., \& Flannery, B.P. 1992, Numerical Recipes in C.~The Art of Scientific Computing (Cambridge: University Press)\\
Thommes, E.W., Duncan, M.J., \& Levison, H.F.~1999, Nature, 402, 635\\
Thommes, E.W., Duncan, M.J., \& Levison, H.F.~2002, \aj, 123, 2862\\
Tiscareno, M.S.~2004, Ph.D.~Thesis, University of Arizona \\
Tiscareno, M.S., \& Malhotra, R.~2003, DPS Meeting \#35, \#39.22 \\
Ward, W.~1997, \apj, 482, L211\\
Winter, O.C., \& Murray, C.D.~1997, A\&A, 328, 399 \\
Wyatt, M.C. 2003, \apj, 598, 1321\\
\end{references}
\end{document}